# IMPROVED ATOMIC TRANSITION PROBABILITIES FOR UV AND OPTICAL LINES OF Hf II AND DETERMINATION OF THE Hf ABUNDANCE IN TWO METAL-POOR STARS[1]

(Short Title: TRANSITION PROBABILITIES OF Hf II)

E. A. Den Hartog[2], J. E. Lawler[2], and I. U. Roederer[3,4]

[2]Department of Physics, University of Wisconsin – Madison, 1150 University Ave, Madison, WI 53706; eadenhar@wisc.edu; jelawler@wisc.edu

[3]Department of Astronomy, University of Michigan, 1085 S. University Ave., Ann Arbor, MI 48109, USA; iur@umich.edu

[4]Joint Institute for Nuclear Astrophysics – Center for the Evolution of the Elements (JINA-CEE), USA

[1] Based on observations made with the NASA/ESA Hubble Space Telescope, obtained at the Space Telescope Science Institute (STScI), which is operated by the Association of Universities for Research in Astronomy, Inc. (AURA) under NASA contract NAS 5-26555. These observations are associated with programs GO-12554, GO-14765, and GO-15657. This paper includes data gathered with the 6.5 meter Magellan Telescopes located at Las Campanas Observatory, Chile.

Abstract

We report new branching fraction measurements for 199 UV and optical transitions of Hf II. These transitions range in wavelength (wavenumber) from 2068 – 6584 Å (48322 – 15183 $cm^{-1}$) and originate in 17 odd-parity upper levels ranging in energy from 38578 – 53227 $cm^{-1}$. The branching fractions are combined with radiative lifetimes reported in an earlier study to produce a set of transition probabilities and log(*gf*) values with accuracy ranging from 5 – 25%. Comparison is made to transition probabilities from the literature where such data exist. We use these new transition probabilities to derive improved Hf abundances in two metal-poor stars. HD 196944 is enhanced in s-process elements, and we derive log $\varepsilon$ (Hf) = -0.72 ± 0.03 ($\sigma$ = 0.09) from 12 Hf II lines. HD 222925 is enhanced in r-process elements, and we derive log $\varepsilon$ (Hf) = 0.32 ± 0.03 ($\sigma$ = 0.11) from 20 Hf II lines. These measurements greatly expand the number of potentially useful Hf II lines for analysis in UV and optical spectra.

# 1. INTRODUCTION

Metal-poor stars record the nucleosynthesis products of the earliest stellar generations. Two dominant production channels for heavy elements have been identified, the rapid (r) and slow (s) processes (see, e.g., reviews by Käppeler et al. 2011 and Cowan et al. 2019). The elemental abundance patterns found in metal-poor stars provide constraints on the physical conditions and astrophysical sites, and large samples of stars in different Galactic environments reveal the changing nature of these production sites across cosmic time (e.g., Côté et al. 2019; Kobayashi et al. 2020).

Stellar abundances must be both precise and accurate to yield reliable information about stellar nucleosynthesis. In metal-poor stars, the strongest and often only detectable lines of heavy elements are found in the blue and ultraviolet (UV) parts of the spectrum (e.g., Roederer et al. 2012). These parts of the spectrum are often crowded, so absorption lines of interest are likely to be blended with lines of other species. To achieve precise abundances, it is critical to derive abundances from many lines of the same species to limit the impact of systematic errors, like unidentified blends. To achieve accurate abundances, accurate atomic data are necessary, and the electronic level populations in the line-forming layers of the atmosphere must be reliably modeled. Some lines may be more reliable abundance indicators than others. These "highly reliable lines" (HRLs) connect upper levels to the ground and low metastable levels of the dominant ion, and they are well-modeled by the standard assumption that local thermodynamic equilibrium (LTE) holds in the line-forming layers.

While analyzing blue and UV spectra of several metal-poor stars with enhanced levels of r- and s-process elements, we noticed many lines of singly-ionized hafnium (Hf, Z = 72), the dominant species of Hf in metal-poor stellar atmospheres. These lines show potential to be useful abundance indicators, including several HRLs, but they lack sufficiently reliable atomic data, most notably absolute transition probabilities. Hf is important from a nucleosynthesis perspective because it links the well-studied Rare Earth Element domain ($57 \leq Z \leq 71$; e.g., Sneden et al. 2009) and the 3rd r-process peak ($76 \leq Z \leq 78$; e.g., Den Hartog et al. 2005). Hf also shows promise as a reliable stable element for use in determining stellar ages from the decay of long-lived radioactive actinide isotopes $^{232}$Th, $^{235}$U, and $^{238}$U (Lawler et al. 2007). The HRLs of Hf II are only found in the spectral range covered by a few blue-sensitive spectrographs on ground-based telescopes and one UV-sensitive spectrograph on board the Hubble Space Telescope (HST). Collecting these spectra places a high demand on competitive, limited community resources, so it is important to maximize the scientific return from each spectrum.

As with most modern studies, our method for determining transition probabilities (Einstein *A*-values) is to combine branching fractions (BFs) determined from high-resolution emission spectra with the radiative lifetime of each upper level, which puts the BFs on an absolute scale. This combination of techniques yields transition probabilities ranging in accuracy from ~5 – 25%. The uncertainty of the strong branches is dominated by the radiative lifetime uncertainty, typically ±5%, while the uncertainty of weak branches is dominated by the uncertainty of the BF. The *A*-values are then converted to log(*gf*)s (the logarithm of the absorption oscillator strength multiplied by the degeneracy of the lower level of the transition) which is a more useful form for astronomers in stellar abundance studies.

Hf II transition probabilities have been the subject of two previous modern experimental efforts. In 2006, the group at Lund University and collaborators reported *A*-values and log(*gf*)s for 195 transitions originating in 18 upper levels of $Hf^+$ for which they also reported lifetimes (Lundqvist et al. 2006, henceforth L06). The following year our University of Wisconsin (UW) group published radiative lifetimes for 41 levels of $Hf^+$ and *A*-values and log(*gf*)s for 150 transitions originating in 19 of these upper levels (Lawler et al. 2007, henceforth UW07). L06 and UW07 overlap for 72 transitions for which generally good agreement was observed, with only a handful of weak lines lying outside the combined uncertainties. While these two studies provide modern, accurate measurements for over 270 transitions between them, there still remain many transitions of Hf II, especially in the deep-UV, that are potentially useful as abundance indicators but have no reliable transition probability data. That is the motivation for the current study. Of the 41 levels with accurate radiative lifetimes reported by UW07, BFs were not measured for 13 in either of the previous two studies. In the case of UW07, this was due to these levels having significant far-UV branches which lay outside the calibration of their spectra. In the current study, we report *A*-values and log(*gf*) values for transitions associated with these 13 levels as well as four of the higher-lying levels previously studied only by L06. The repetition of a few of the levels studied by L06 is to ascertain that no significant discrepancies are observed for levels requiring a calibration extending deep into the UV.

## 2. BRANCHING FRACTIONS OF Hf II

The BF for a transition between an upper level *u* and lower level *l* is the ratio of its *A*-value to the sum of all the *A*-values associated with *u*. This can also be expressed as the ratio of relative emission intensities I (in any units proportional to photons/time) for these transitions:

$$BF_{ul} = \frac{A_{ul}}{\sum_l A_{ul}} = \frac{I_{ul}}{\sum_l I_{ul}}. \quad (1)$$

BFs, by definition, sum to one. It is therefore important when measuring BFs to account for all possible decay paths from an upper level so that the normalization is correct. The energy level structure of $Hf^+$ is discussed in some detail in the UW07 paper. Suffice it to say here that the configurations of both parities are sufficiently well known and there are few missing low-lying levels of even parity that might give rise to missing branches in the current study. All possible transitions which obey the ΔJ and parity change selection rules are investigated. We use the energy levels from Moore (1971) which are downloaded from the National Institute of Science and Technology Atomic Spectra Database (NIST ASD)[5] (Kramida et al. 2020). Transition wavenumbers are calculated from the difference in level energies and these are converted to transition wavelengths using the standard index of air (Peck & Reeder 1972). We also made detailed comparison of our set of transitions for each upper level to the classified line list of Meggers & Scribner (1934). The L06 paper reported a set of improved wavelength measurements for lines of Hf II, but careful comparison to our spectra indicate that some of the wavelength measurements may have problems. L06 did not report a comprehensive set of improved energy levels. We therefore choose to use the older self-consistent set of levels from Moore in this study. Research is continuing at UW on improved energy levels from Fourier transform spectrometer (FTS) data.

---

[5] Available at https://www.nist.gov/pml/atomic-spectra-database

This study in Hf II utilizes two different high-resolution spectrometers, a FTS that covers the near-UV to IR and the UW 3-m high resolution echelle spectrograph for the far-UV to near-UV. Branching ratios (BRs) are determined for the transitions covered by each instrument with several lines in the near-UV determined using spectra from both spectrometers. These lines are then used as a bridge to combine the two sets of BRs, which are then renormalized to determine BFs. These are then combined with lifetimes reported by our group in the earlier UW07 study to determine *A*-values.

As in the UW07 study, line intensities for branches in the near-UV through near-IR spectral regions were measured from emission spectra of custom water-cooled Hf hollow cathode lamps (HCLs) recorded with the 1-meter FTS on the McMath telescope at the National Solar Observatory (NSO), Kitt Peak, AZ. Regrettably, this instrument has since been decommissioned, but all spectra recorded on it are available for download from the NSO archives.[6] The Hf FTS spectra utilized in the current study were recorded in the 1980's and are the same spectra used in the UW07 study. The reader is referred to Table 2 of that publication for details of these spectra. Of the spectra listed, spectra with index numbers 1-5 were the most useful in this study. The 1-m FTS has many advantages for branching fraction work and has been the work horse of the UW transition probability effort for decades. It has broad spectral coverage from the UV to IR, a limit of resolution down to 0.01 $cm^{-1}$, and absolute wavenumber accuracy of 1 part in $10^8$. As an interferometric instrument it has large etendue and is insensitive to small drifts in source current, as data is collected on all spectral elements simultaneously. For BF work, it is important to establish an accurate relative radiometric calibration. The calibration method of choice is to use the Ar I & II BR technique which employs well known BRs in the range 2857 – 23256 Å (35,000 – 4300 $cm^{-1}$) established for this purpose by Adams & Whaling (1981), Danzmann & Kock (1982), Hashiguchi & Hasikuni (1985), and Whaling et al. (1993). The Ar I & II BRs are determined in exactly the same manner as the BRs of the specie under study, in this case Hf II. Because the technique is totally internal to the measurement, it captures effects such as window transmission and reflections within the lamp that a calibration using an external standard lamp could miss. The analysis of the FTS spectra was carried out interactively, with baselines and integration limits set by hand and a numerical integral made of the line intensity. The intensities for the Ar I & II lines are compared to the known BRs to construct a sensitivity curve for each spectrum, which is then used to calibrate the Hf II intensities.

One drawback of any FTS is that it is not ideal for measuring the intensity of weak lines, as the Poisson statistical noise of all lines in the spectrum is spread evenly throughout the spectrum. This is known as multiplex noise, and results in the weak lines being swamped in the noise from the strong lines in the spectrum. As the source current is increased to bring the weak branches above the noise floor the strong branches are increasingly affected by optical depth. This situation affects the BFs, making the weak branches appear stronger than they would in a spectrum from an optically thin source. Corrections based on line shape are sometimes attempted but are difficult and have an adverse effect on the accuracy of the BFs. The desire to measure weaker branches in the UV while keeping source currents modest led our group to develop the UW high resolution 3-m echelle spectrometer (Wood & Lawler 2012). This instrument has high resolving power (>250,000), broad spectral coverage and high UV sensitivity. As a dispersive instrument it is free of multiplex noise and can be used to measure weak lines with much better signal-to-noise (S/N) than an FTS. It does not, however, have as good an absolute wavenumber accuracy as an FTS.

[6] FTS data are publicly available at http://diglib.nso.edu/

The UW 3-m echelle spectrograph was used to measure the Hf II BRs in the deep- to near-UV. New spectra were recorded for this study using a custom see-through water-cooled demountable hollow cathode lamp. These spectra are 2-dimensional CCD images of multiple grating orders with the high-resolution of a given grating order running in one direction, and the grating orders arrayed side-by-side in the other direction. In the UV it takes three overlapping CCD frames to capture the full breadth of the grating order. In most of our spectra we use five frames for some redundancy and to check for effects of source drift. Each frame is calibrated with a spectrum of a NIST traceable deuterium ($D_2$) standard lamp recorded immediately after the HCL spectra. The only change made to the instrument between these two measurements is the rotation of a steering mirror on a kinematic mount. This "everyday" $D_2$ lamp is also periodically checked against a little-used NIST traceable $D_2$ lamp to correct for lamp aging. The calibration of the $D_2$ lamp contributes <2% to the uncertainty of the relative line intensities. Because the calibration of the 3-m echelle uses an external source, it is important to know and correct for changes in the HCL window transmission as function of wavelength. The window transmission was measured by recording a $D_2$ lamp spectrum with and without the HCL in the light path. This measurement was made with the cathode removed from the lamp so that the $D_2$ lamp light only encountered the front and rear window of the HCL body.

Table 1 summarizes the 3-m echelle spectra recorded for this study. They range in current from 10 – 50 mA and use either ~5 Torr Ne or ~2 Torr Ar as the buffer gas. The first 25 spectra are five groups of five CCD frames. Each group of five is a full spectrum over the listed spectral range. The next spectrum listed in Table 1 (index number 46) is a single far-UV frame to check for two possible transitions that fell outside the spectral ranges of the first 25 spectra. The first is the 1992.32 Å line from the $w^2F^o_{7/2}$ upper level at 53227.27 $cm^{-1}$. This line was observed and yielded a BF=0.004, but with high uncertainty because its wavelength is slightly outside the range of the radiometric calibration. This line (as with several other weak lines with high uncertainty) is dropped from our final table but retained in the BF normalization. The other possible far-UV line is at 2039.93 Å from the $x^2D^o_{5/2}$ upper level 49005.64 $cm^{-1}$. This line was not observed. The final seven spectra at the bottom of Table 1 are individual CCD frames that were used primarily to study specific line pairs to resolve blending issues, as discussed below. Like the FTS spectra, the echelle spectra are analyzed interactively. Baselines and integration limits are set by hand and a numerical integral is made of the relative line intensity. In the case of the echelle spectrum a second integral is made of the $D_2$ lamp spectrum at the same location on the CCD. The $D_2$ spectrum captures both the gradual change of instrument sensitivity from one order to the next as well as the $sinc^2$ intensity envelope from the grating along the high-resolution direction of each order.

If the sum in the denominator of Equation 1 is over some but not all possible lower levels, then one has an un-normalized BR. The BRs for the transitions in the far-UV up to ~3800 Å (26,300 $cm^{-1}$) are determined from the relative intensities as measured from the 3-m echelle spectra. BRs for the transitions from ~2850 Å (35,000 $cm^{-1}$) to the near-IR are determined from the FTS spectra. These two sets of BRs are put on the same scale using lines that they have in common in the near-UV for each upper level. For the levels in this study these bridge lines number from a minimum of two up to seven with most levels having four or five bridge lines. Once the BRs are put on the same relative scale, they are renormalized as BFs to sum to one.

The uncertainties on final mean BFs are evaluated from the strength of the BFs, the S/N of the spectral lines, and the wavenumber difference of lines from the common upper level. By definition, branching fractions from a given upper level sum to 1.00, so uncertainty migrates to the weaker lines. The relative radiometric calibration also contributes to the uncertainty. The conservative estimate of the calibration uncertainty is 0.001%/cm$^{-1}$ of the wavenumber difference between a line and the dominant branch(es) from the common upper level (Wickliffe et al. 2000).

### 2.1 Blends

Although not as rich as the nearby rare earth elements, a Hf HCL spectrum has a high density of lines and the potential for line blends must be considered. The analysis software used in this work automatically looks for possible blends from Hf I, Hf II and from the first and second spectrum of the buffer gas, either Ne or Ar. In this study we saw an unusually high number of unidentified lines that, upon detailed investigation, turned out to be from the first or second spectrum of Zr. In retrospect, this is not too surprising because Hf is always found in nature in Zr minerals such as zircon, with Hf replacing 1 - 4% of the Zr. Furthermore, Hf is very difficult to separate from Zr due to their similar chemistries. So, in addition to considering the usual blending suspects we searched the MIT Wavelength Tables (Phelps III 1982) line list of Zr I & II for exact or near potential blends. We found five blends with lines of Zr which are not resolved in our spectra. Fortunately, three of these lines were included in a transition probability study of Zr II by Ljung et al. (2006). We were able to measure intensities of nearby Zr II lines from the same upper levels and, using the BFs from that study, calculate the intensity of the blending line and subtract it from our line integrals. These blends are:

- 2712.42 Å: The $a^4F_{5/2}$ - $y^4D_{5/2}$ transition of Zr II blended with the $a^4F_{5/2}$ - $y^2D_{5/2}$ transition of Hf II. The Zr II line is ~2% of the total line intensity.
- 2968.96 Å: the $b^4F_{9/2}$ - $y^4F_{7/2}$ transition of Zr II blended with the $a^4P_{5/2}$ – $z^2G_{7/2}$ transition of Hf II. The Zr II line is ~17% of the total line intensity.
- 2898.71 Å: The $b^4F_{7/2}$ - $y^2F_{7/2}$ transition of Zr II blended with the $a^4P_{5/2}$ – $x^2D_{3/2}$ transition of Hf II. The Zr II line is ~1% of the total line intensity.

The other two unresolved blends are 2400.814 Å line of Zr blending with 2400.82 Å from the $y^2F^o_{7/2}$ level of Hf II and 2467.971 Å line of Zr II blending with 2467.97 Å from the $z^4P^o_{5/2}$ level of Hf II. These blends are expected to be insignificant as they are listed in the MIT Tables with logarithmic spark intensities of 0.0 and 0.3, respectively, while the three blends detailed above have logarithmic spark intensities of 1.2, 1.5 and 0.8, respectively.

Broadening and structure of the Hf II lines also make sorting out blends more difficult, particularly when the blend is suspected but the blending partner is unknown. Hafnium has six stable isotopes (*A* = 174, 176, 177, 178, 179, 180), five of which have non-negligible isotopic fraction (*A*=174 has only 0.16% in a naturally occurring isotopic mix). One might therefore expect to see evidence of isotopic structure, and two of these isotopes are odd and may therefore also exhibit hyperfine structure. We do observe broadening and asymmetry in many lines of our spectra, although the isotopic and hyperfine structure is not resolved. Figure 1 shows an example of such broadening in a Hf II line as observed in one of the 3-m echelle spectra. The broad pedestal, probably due to hyperfine structure, is not uncommon in the observed line shapes. Also seen in this figure is a resolved contaminant line identified as belonging to Zr II. Our search through the MIT Wavelength Tables (Phelps III 1982) also helped us identify several near blends as Zr I & II that we originally suspected of being isotopic or hyperfine structure. We were then able to eliminate this nearby structure from the integrated line intensity.

Two other blends need some mention here. The first is the line 3979.38 Å (25122.44 $cm^{-1}$) from upper level near 53227 $cm^{-1}$ which is blended with an Ar II line. The usual way to handle a buffer gas line blend, is to look at the line in a spectrum using the other buffer gas, in this case Ne. However, the two Ne FTS spectra listed in Table 2 of the UW06 paper and consulted in the current study had inadequate S/N to be useful. The blend was therefore separated in the Hf-Ar FTS spectra using the center-of-gravity (COG) technique which is described in Lawler et al. (2018) and in other publications. Lawler et al. state, "The COG of the blended feature is compared to Ritz wave numbers for both contributing transitions. Then a requirement is imposed that a normalized and weighted combination of the Ritz wave numbers match the COG of the blended feature. This COG method yields the fractional contribution of each line to the blended feature. For good reliability the COG technique requires a high S/N ratio as well as a Ritz wave number separation of $\approx$ 0.05 $cm^{-1}$ or more and an accurate wave number calibration of the FTS data." The COG technique also requires accurate energy levels to compute accurate Ritz wavenumbers for the blended transitions. The Ar II energy levels of NIST ASD are reported to 0.0001 $cm^{-1}$, but the Hf II energy levels of NIST ASD are reported to 0.01 $cm^{-1}$. Generally uncertainties are assumed to be a few times the last digit of the energy level. The Ar II Ritz wavenumbers are thus accurate to ~0.0003 $cm^{-1}$ which is fine for the COG technique, but the Hf II Ritz wavenumbers are accurate to ~0.03 $cm^{-1}$ which is not quite satisfactory for COG technique. L06 reported improved Hf II energy levels for several lower levels of lines from the upper level near 53227 $cm^{-1}$ including the lower level of the 25122 $cm^{-1}$ line and the lower level of the 30082 $cm^{-1}$ line. We used the 30082 $cm^{-1}$ line to determine an improved upper level energy of 53227.359 $cm^{-1}$ compared to the NIST ASD value of 53227.27 $cm^{-1}$. Then we used the COG separation technique on the 25122 $cm^{-1}$ blend with the improved lower level energy from L06 and our improved upper level energy. The BF from the COG technique was compared to, and averaged with, a BF measured using a Ne-Hf HCL and the 3-m echelle. Three single frame spectra were recorded with the echelle to determine the BR of this transition relative to branches at 2814.76 and 3323.32 Å (35516.55 and 30081.70 $cm^{-1}$). These are the spectra in Table 1 with index numbers 48 – 50. The weighted average BF = 0.124 lies within the respective estimated uncertainties arising from the FTS COG separation and the 3-m echelle measurement.

A more difficult blend to resolve is that of the 2898.71 Å line belonging to upper level $x^2D_{3/2}$ near 47974 $cm^{-1}$. As discussed above, this transition has a weak Zr II blend which accounts for only ~1% of the total line intensity, but there appears to be another unknown and unresolved blend. This transition shows markedly different intensity ratio with the transition from the same upper level at 2852.02 Å in Ar buffer gas compared to Ne. Furthermore, the ratio in Ne was constant at ~1.3 over multiple spectra taken over a factor of five range in current. The same ratio ranged from 1.7 – 2.7 in different Ar spectra. We first suspected the blending partner was an Ar

I or II line because the ratio is both higher and non-constant in Ar compared to Ne. However, no such line appears in available Ar I & II line lists. To be absolutely certain that the unknown blend did not belong to Ar I or II we checked archived 3-m echelle spectra of Co-Ar HCLs to see if a line showed up at 2898.71 Å, but none was found. Having eliminated an Ar line as the potential blending partner, we took several single-frame spectra that had both the blended line and reference line on the frame. The first was of a Hf - Ne commercial HCL running at 10 mA (spectrum 51 in Table 1). The thought was that if the blend was from a contaminant in our Hf sample the commercial lamp might have a different purity and therefore give a different ratio. This spectrum gave a ratio consistent with those measured in our custom HCL running Ne as buffer. We then measured three single frame spectra of the custom HCL in Ar buffer to expand the range of HCL currents to a factor of eight (spectra 52 – 54 in Table 1). No clear trend with current was observed, but the ratio in Ar varied widely. Although we would like to have identified the blend definitively, our conclusion is that whatever the blend is, it is significant in our Ar-Hf HCL spectra and insignificant in our Ne-Hf HCL spectra. We draw this conclusion because of the stability of the ratio in Ne over a wide range of conditions. We therefore have determined the BF for the 2898.71 Å line strictly from the Ne 3-m echelle spectra, but with increased uncertainty due to the potential of residual blend.

## 3. RESULTS AND COMPARISON TO EARLIER STUDIES

Our BFs from this study are presented in Table 2 organized by term-ordered upper level. Also in this table we list BFs of L06 for the levels that are in common. L06 only quote uncertainties for their *gf* values, which include both the BF uncertainties and lifetime uncertainties. To better approximate their BF uncertainties in Table 2 we quote their total uncertainties reduced in quadrature by their lifetime uncertainties. In Figure 2 we compare our BF results to L06 for the transitions in common by plotting the ratio $BF_{L06}/BF_{UW}$ versus $BF_{UW}$ in the top panel and versus wavelength in the bottom panel. The level of agreement is generally satisfactory, with just a few lines lying beyond their combined uncertainties from the horizontal line at 1.0 that represents perfect agreement.

Our BFs are converted to transition probabilities and subsequently to log(*gf*)s using the relations from Thorne (1988), where $A_{ul}$ is the transition probability in $s^{-1}$, $\tau_u$ is the radiative lifetime of the upper level in s, $g_u$ is the degeneracy of the upper level, and $\sigma$ is the transition wavenumber in $cm^{-1}$:

$$A_{ul} = \frac{BF_{ul}}{\tau_u} \quad ; \quad \log(gf) = \log(1.499 g_u A_{ul}/\sigma^2) \tag{2}$$

We use our lifetimes from the UW07 study in the conversion to *A*-values. We present our *A*-values and log(*gf*)s for 199 transitions of Hf II in Table 3 which is ordered by increasing wavelength. This table is available as a machine-readable ascii table in the online publication. We make one final comparison of our results to the theoretical log(*gf*)s of Bouazza, Quinet & Palmeri (2015) who used the Hartree-Fock model including core polarization effects to calculate log(*gf*)s for approximately 1000 transitions of Hf II (see their spreadsheet of results in the appendix of that paper). Our comparison for the transitions which overlap in our studies is shown in Figure 3. The top panel includes a comparison for all transitions from upper levels in common. The level of agreement is not particularly good. In the bottom panel we remove the transition from four upper levels, $y^4F^o_{5/2}$, $y^4F^o_{7/2}$, $y^2F^o_{5/2}$ and $y^2F^o_{7/2}$, from the comparison. The most severe outliers belong to these levels and the agreement for all remaining levels is much more satisfactory. Nearby levels with same parity and J values mix due to a breakdown in Russell-Saunders (LS) coupling which may explain why these levels are more difficult to handle theoretically.

## 4. APPLICATION TO TWO METAL-POOR STARS

To demonstrate the utility of these new measurements for stellar astrophysics, we apply them to the analysis of spectra of two stars that are highly enhanced in heavy elements. One, HD 196944, is a carbon-enhanced metal-poor star with an excess of s-process elements (e.g., Začs et al. 1998; Van Eck et al. 2001). The other, HD 222925, is a metal-poor star with an excess of r-process elements (Roederer et al. 2018).

We gather high-quality UV and optical spectra of these stars from public archives and our own observations. The UV spectrum of HD 196944 was obtained during two sets of observations using the Space Telescope Imaging Spectrograph (STIS; Kimble et al. 1998; Woodgate et al. 1998) on board HST. One spectrum covers 2029 – 2303 Å with resolving power, R, = 114,000 (GO-14765, datasets OD5A01010-14010, PI Roederer), and the other covers 2278 – 3073 Å with R = 30,000 (GO-12554, datasets OBQ601010-30, PI Beers). The UV spectrum of HD 222925 was also obtained with STIS, and it covers 1936 – 3145 Å with R = 114,000 (GO-15657, datasets ODX901010-60030, PI Roederer). These spectra were downloaded from the Mikulski Archive for Space Telescopes (MAST) and processed automatically by the CALSTIS software package. The optical spectrum of HD 196944 was obtained using the Magellan Inamori Kyocera Echelle (MIKE) spectrograph (Bernstein et al. 2003) on the Magellan II (Clay) telescope at Las Campanas Observatory, Chile. It covers 3350 – 9150 Å at R = 41,000 ($\lambda < 5000$ Å) and R = 35,000 ($\lambda > 5000$ Å), as described in more detail in Roederer et al. (2014). The optical spectrum of HD 222925 was also obtained with MIKE, and it covers 3330 – 9410 Å at R = 68,000 ($\lambda < 5000$ Å) and R = 61,000 ($\lambda > 5000$ Å), as described in more detail in Roederer et al. (2018).

We inspect these spectra at the wavelengths of the Hf II lines listed in Table 3. Most are too weak to be detected, and many of the stronger lines are prohibitively blended with lines of other species, most commonly Fe I or Fe II. For all potentially useful Hf II lines, we synthesize a 6-Å region of the spectrum using a recent version of the LTE line analysis code MOOG (Sneden 1973; Sobeck et al. 2011). We generate line lists for all potential blending features using an updated version of the linemake code[7]. We interpolate model atmospheres from the ATLAS9 grid of α-enhanced models (Castelli & Kurucz 2004). For HD 196944, we use the model parameters derived by Placco et al. (2015): effective temperature ($T_{eff}$) = 5170 K, log of the surface gravity (log $g$) = 1.60, microturbulent velocity ($v_t$) = 1.55 km $s^{-1}$, model metallicity ([M/H]) = -2.41, and iron abundance ([Fe/H]) = -2.41. For HD 222925, we use the model parameters derived by Roederer et al. (2018): $T_{eff}$ = 5636 K, log $g$ = 2.54, $v_t$ = 2.20 km $s^{-1}$, [M/H] = -1.50, and [Fe/H] = -1.46.

[7] https://github.com/vmplacco/linemake

We derive Hf abundances from 12 Hf II lines in HD 196944 and 20 Hf II lines in HD 222925, including Hf II lines whose transition probabilities were presented in the UW07 study. These results are shown in Table 4. Previously, only 1 and 5 lines of Hf II, respectively, had been used as abundance indicators in these stars, although results from analyses of parts of the UV spectra of these stars had not previously been reported in the literature. Figure 4 illustrates our synthetic spectra fits to a few of the lines in each of these stars. Although some blending features are present, the Hf II lines are clearly detected and yield reliable, consistent abundances.

The weighted mean Hf abundances in HD 196944 and HD 222925 are log ε (Hf) = -0.72 ± 0.03 (σ = 0.09) and log ε (Hf) = 0.32 ± 0.03 (σ = 0.11), yielding [Hf/Fe] = +0.84 and +0.92, respectively. The uncertainties account for uncertainties in the transition probabilities (Table 3) and uncertainties in the spectrum fitting (Table 4). Uncertainties arising from errors in the model atmosphere parameters are comparable to the standard deviations, σ, reported above (e.g., Roederer et al. 2018).

The log ε (Hf/Eu) ratio is a common way to express the purity of r-process material in a star. We calculate log ε (Hf/Eu) = -0.06 ± 0.08 in HD 222925 by combining our Hf abundance with the Eu abundance derived by Roederer et al. (2018). This value is in excellent agreement with modern assessments of the log ε (Hf/Eu) ratio found in r-process material in the solar system, -0.05 (from both Sneden et al. 2008 and Bisterzo et al. 2011), indicating a pure r-process origin with minimal contributions from the s-process.

## 5. SUMMARY

We report new branching fraction measurements for 199 UV and optical transitions of Hf II originating in 17 odd-parity upper levels ranging in energy from 38578 – 53227 cm$^{-1}$. The branching fractions are combined with radiative lifetimes reported in an earlier study to produce a set of transition probabilities and log(*gf*) values with accuracy ranging from 5 – 25%. Comparison is made to transition probabilities from experimental and theoretical data from the literature.

We also apply our new set of Hf II transition probabilities to high-resolution UV (HST/STIS) and optical (Magellan/MIKE) spectra of two metal-poor stars, one that is enhanced in r-process elements and another that is enhanced in s-process elements. Our results demonstrate the utility of the new laboratory measurements to yield both precise and accurate abundances when applied to high-quality spectra of metal-poor stars, and the new transition probabilities greatly expand the number of potentially useful Hf II lines for stellar abundance analyses.

## ACKNOWLEDGEMENTS

This work is supported by NSF grant AST1814512 (E.D.H. and J.E.L) and all authors acknowledge support from NASA through grants GO-14765 and GO-15657 from STScI, which is operated by the AURA under NASA contract NAS5-26555. I.U.R. also acknowledges support from grants PHY 14-30152 (Physics Frontier Center/JINA-CEE) and AST-1815403, awarded by the U.S. National Science Foundation. This research has made use of NASA's Astrophysics Data System Bibliographic Services, the arXiv pre-print server operated by Cornell University, the ASD hosted by NIST, the MAST at STScI, and the Image Reduction and Analysis Facility (IRAF) software package.

Facilities: HST (STIS), Magellan (MIKE).

Software: IRAF (Tody et al. 1993), linemake, matplotlib (Hunter 2007), MOOG (Sneden 1973), numpy (van der Walt et al. 2011), R (R Core Team 2013).

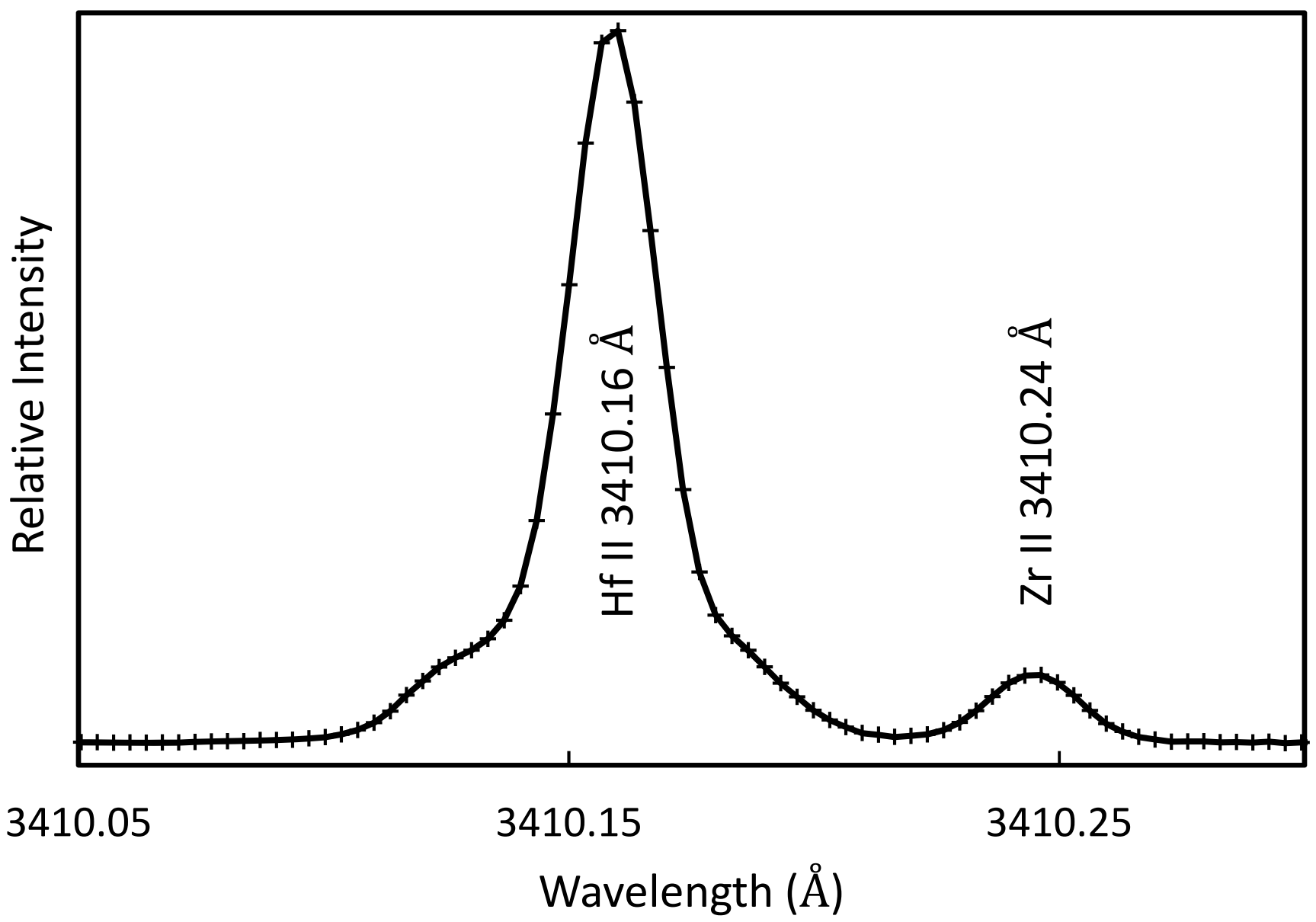

Figure 1. Example from a 3-m echelle spectrum of the observed broadening of a Hf II line due to unresolved isotopic and hyperfine structure. The + symbols indicate the relative number of counts in each pixel. The solid line simply connects these symbols as an aid to the eye. The broad pedastal is typical of many Hf II lines. Also shown is a nearby Zr II contaminant line.

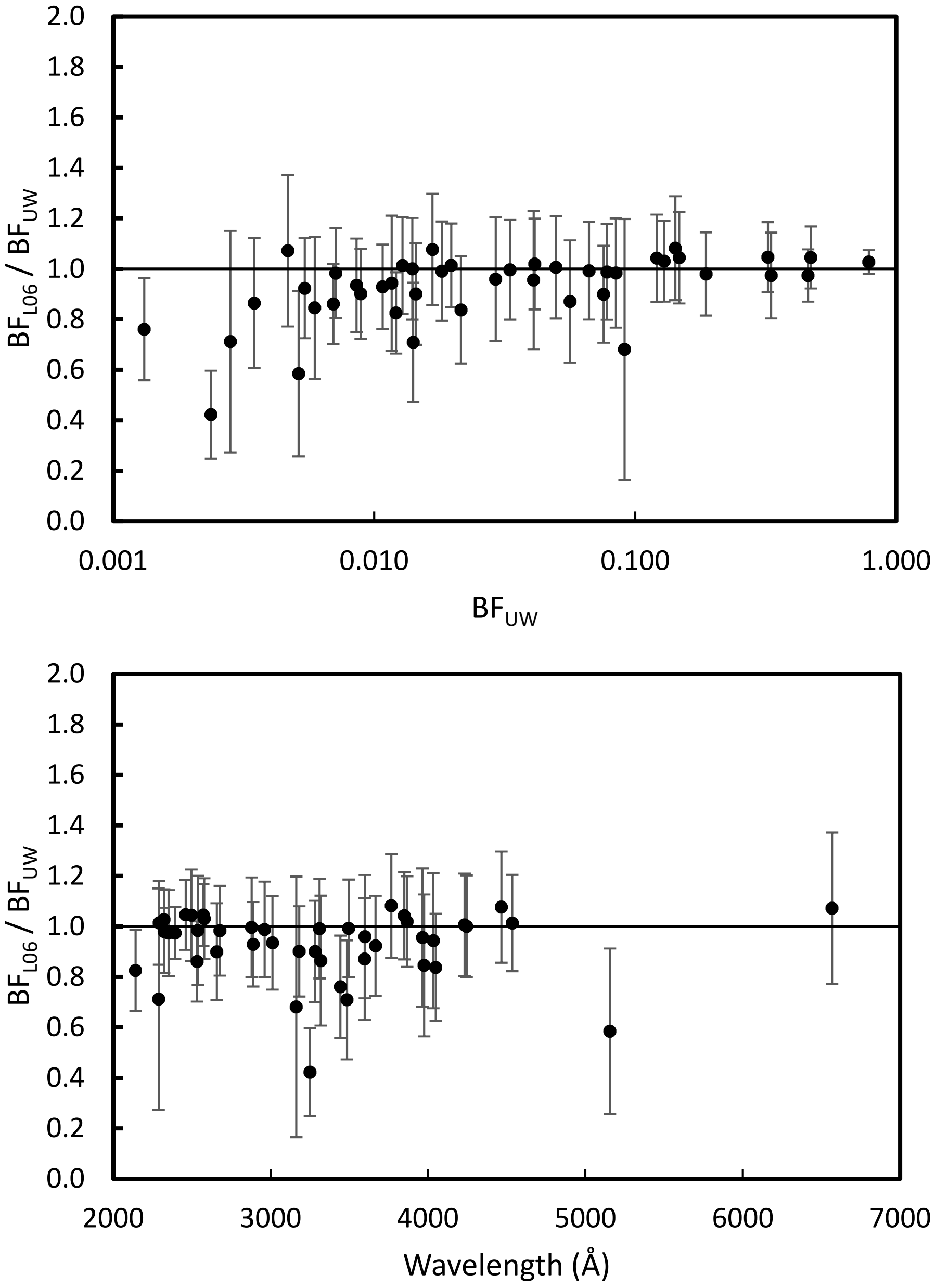

Figure 2. Ratio of experimental BFs from L06 to those of this study versus BFs from this study (upper panel) and versus wavelength (lower panel). The solid horizontal line represents perfect agreement. Error bars represent the uncertainties of the BFs combined in quadrature.

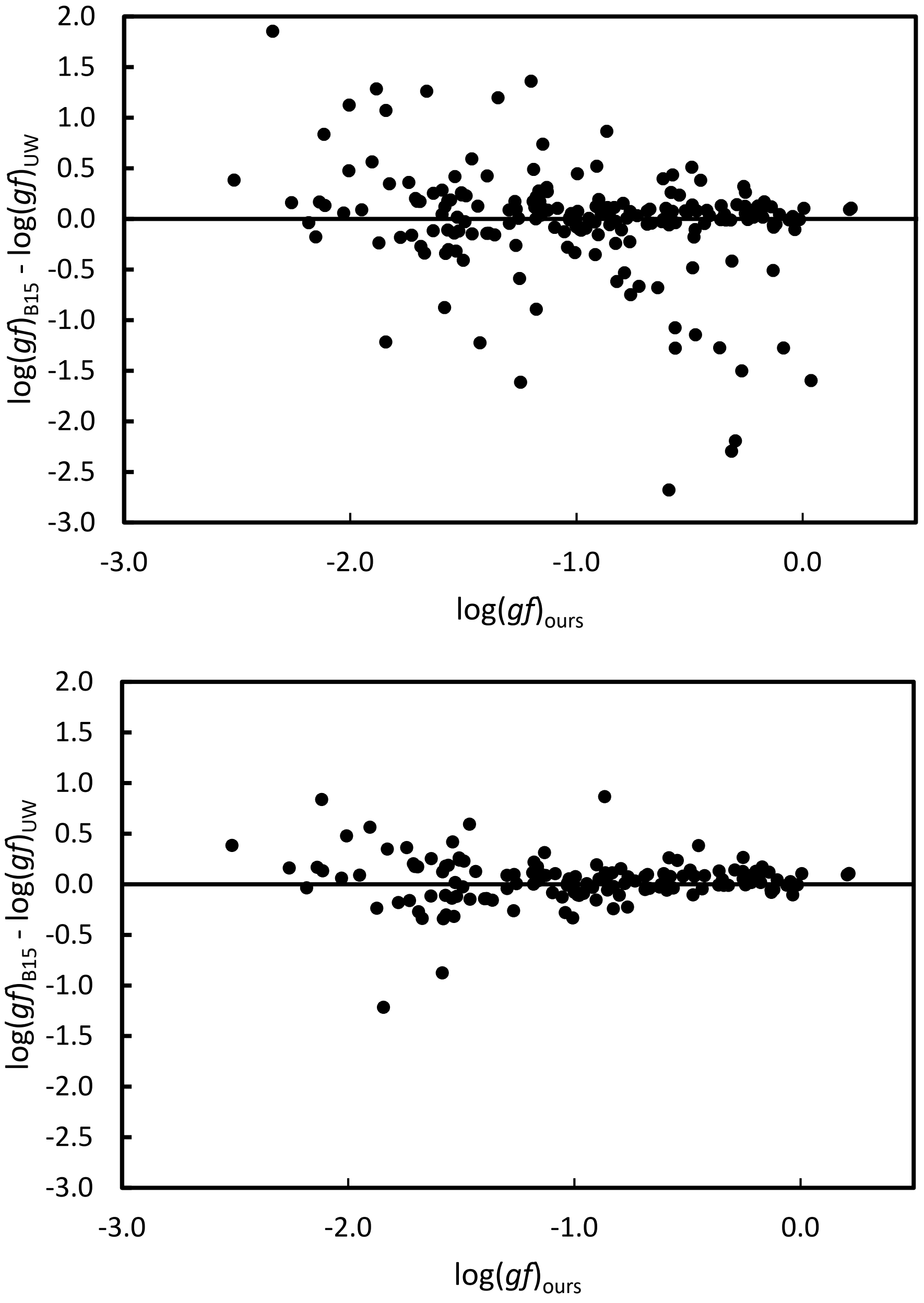

Figure 3. Comparison of the log(*gf*)s from Bouazza et al. (2015) to those from this study. The top panel is a comparison for transitions from all upper levels in this study. The lower panel has the transitions from four levels removed from the comparison. These are $y^4F^o_{5/2}$, $y^4F^o_{7/2}$, $y^2F^o_{5/2}$ and $y^2F^o_{7/2}$. It can be seen that most of the severe outliers in the comparison came from these four levels.

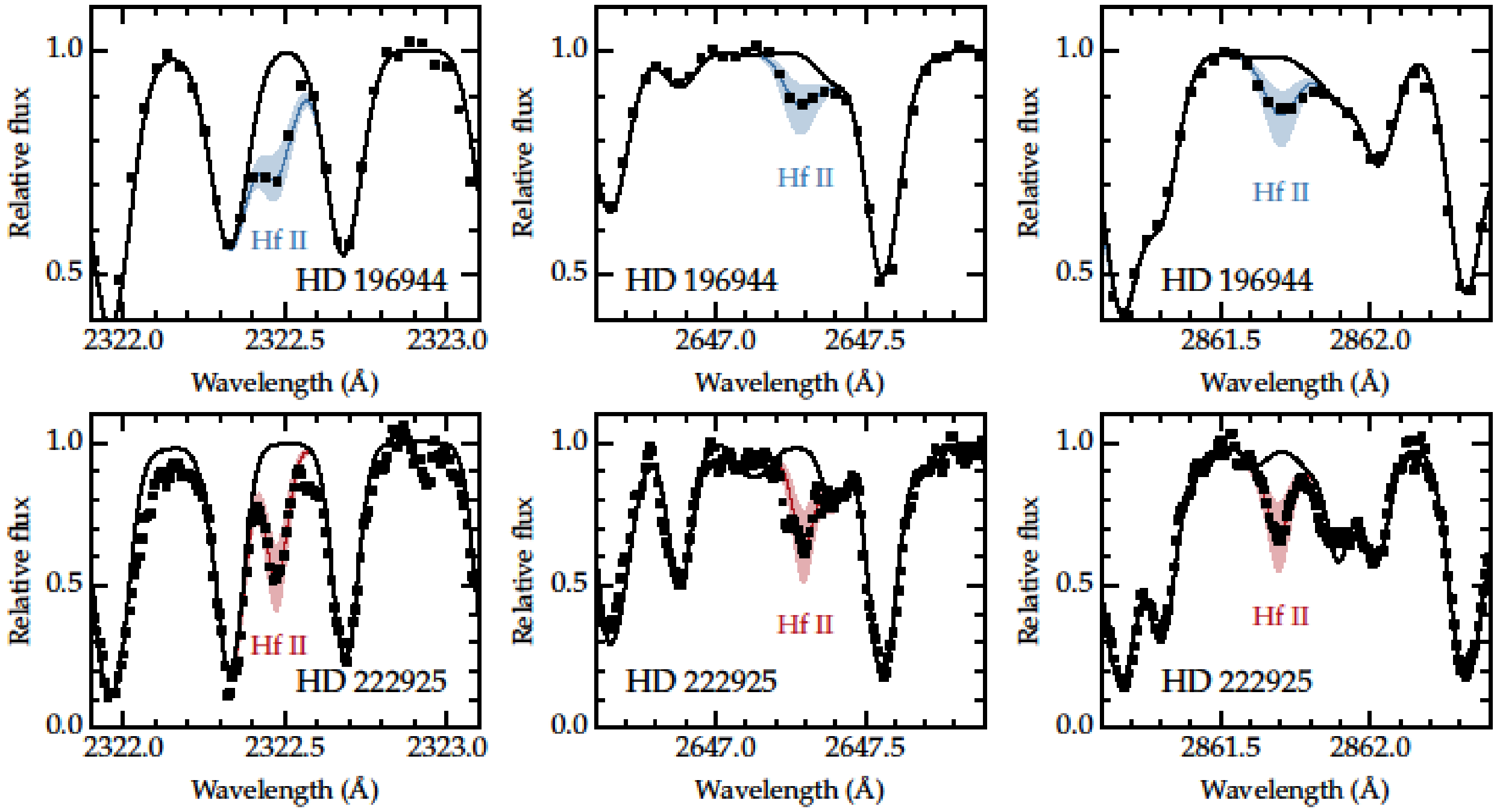

Figure 4. Comparison of observed and synthetic spectra for three Hf II lines in each of HD 196944 and HD 222925. The blue or red solid lines (blue for HD 196944, red for HD 222925) represent synthetic spectra with the best-fit Hf abundance, and the shaded bands represent changes in this abundance by factors of ± 2 (0.3 dex). The black lines represent synthetic spectra with no Hf. The points mark the observed spectra. The higher spectral resolution of the HD 222925 spectrum, relative to the HD 196944 spectrum, is evident.

Table 1. Echelle spectra of custom demountable see-through HCL.[a]

| Index[b] | Date | Serial Number | Buffer Gas | Lamp Current (mA) | Spectral Coverage (Å) | Coadds | Total Exposure (min) |
|---|---|---|---|---|---|---|---|
| 21 | 2020 Aug 20 | 1 | Neon | 30 | 2080 - 3000 | 120 | 40 |
| 22 | 2020 Aug 20 | 3 | Neon | 30 | 2080 - 3000 | 27 | 40 |
| 23 | 2020 Aug 20 | 5 | Neon | 30 | 2080 - 3000 | 133 | 40 |
| 24 | 2020 Aug 20 | 7 | Neon | 30 | 2080 - 3000 | 96 | 40 |
| 25 | 2020 Aug 20 | 9 | Neon | 30 | 2080 - 3000 | 120 | 40 |
| 26 | 2020 Aug 27 | 1 | Neon | 15 | 2080 - 3000 | 80 | 40 |
| 27 | 2020 Aug 27 | 3 | Neon | 15 | 2080 - 3000 | 27 | 40 |
| 28 | 2020 Aug 27 | 5 | Neon | 15 | 2080 - 3000 | 80 | 40 |
| 29 | 2020 Aug 27 | 7 | Neon | 15 | 2080 - 3000 | 120 | 90 |
| 30 | 2020 Aug 27 | 9 | Neon | 15 | 2080 - 3000 | 120 | 60 |
| 31 | 2020 Aug 28 | 1 | Argon | 20 | 2080 - 3000 | 80 | 40 |
| 32 | 2020 Aug 28 | 3 | Argon | 20 | 2080 - 3000 | 160 | 80 |
| 33 | 2020 Aug 28 | 5 | Argon | 20 | 2080 - 3000 | 164 | 60 |
| 34 | 2020 Aug 28 | 7 | Argon | 20 | 2080 - 3000 | 171 | 80 |
| 35 | 2020 Aug 28 | 9 | Argon | 20 | 2080 - 3000 | 120 | 40 |
| 36 | 2020 Aug 29 | 1 | Argon | 10 | 2200 - 3800 | 180 | 30 |
| 37 | 2020 Aug 29 | 3 | Argon | 10 | 2200 - 3800 | 80 | 60 |
| 38 | 2020 Aug 29 | 5 | Argon | 10 | 2200 - 3800 | 327 | 60 |
| 39 | 2020 Aug 29 | 7 | Argon | 10 | 2200 - 3800 | 200 | 60 |
| 40 | 2020 Aug 29 | 9 | Argon | 10 | 2200 - 3800 | 360 | 60 |
| 41 | 2020 Aug 31 | 1 | Neon | 50 | 2200 - 3800 | 200 | 27 |
| 42 | 2020 Aug 31 | 3 | Neon | 50 | 2200 - 3800 | 100 | 18 |
| 43 | 2020 Aug 31 | 5 | Neon | 50 | 2200 - 3800 | 90 | 14 |
| 44 | 2020 Aug 31 | 7 | Neon | 50 | 2200 - 3800 | 100 | 18 |
| 45 | 2020 Aug 31 | 9 | Neon | 50 | 2200 - 3800 | 90 | 14 |
| 46 | 2020 Nov 07 | 1 | Argon | 20 | 1900 - 2500 | 20 | 60 |
| 47 | 2020 Nov 14 | 1 | Argon | 20 | 2410 - 4000 | 300 | 40 |
| 48 | 2020 Nov 14 | 3 | Neon | 30 | 2410 - 4000 | 400 | 40 |
| 49 | 2020 Nov 21 | 1 | Neon | 30 | 2410 - 4000 | 720 | 30 |
| 50 | 2020 Nov 21 | 3 | Neon | 50 | 2410 - 4000 | 1500 | 30 |
| 51 | 2020 Dec 12 | 1 | Neon | 10 | 2410 - 4000 | 277 | 60 |
| 52 | 2020 Dec 19 | 1 | Argon | 50 | 2410 - 4000 | 300 | 20 |
| 53 | 2020 Dec 19 | 3 | Argon | 35 | 2410 - 4000 | 200 | 20 |
| 54 | 2020 Dec 19 | 5 | Argon | 6 | 2410 - 4000 | 80 | 40 |

Notes:
[a]All echelle spectra were taken of a custom see-through demountable HCL except for spectrum 51 which was taken of a commercial Hf-Ne HCL. All echelle spectra have a spectral resolving power of ~250,000 although the effective resolving power is somewhat lower due to line broadening. Three CCD frames are needed to capture a complete echelle grating order in the UV. Five CCD frames are used for each lamp operating condition for spectra 20 - 45 to provide redundancy and a check for lamp drift. Each of the spectra were calibrated with a $D_2$ lamp spectrum, which was recorded immediately following the completion of each HCL spectrum.
[b]The final eight single frame spectra do not cover an entire echelle grating order. Each of these spectra were recorded to determine a branching ratio for a specific pair of transitions to resolve blends or other questions.

Table 2. Branching Fractions in Hf II

| Upper level[a] | | Lower level[a] | | $\lambda_{air}$ | $\sigma_{vac}$ | Branching Fractions | | | |
|---|---|---|---|---|---|---|---|---|---|
| term | $E_k$ (cm$^{-1}$) | term | $E_i$ (cm$^{-1}$) | (Å) | (cm$^{-1}$) | This expt. | (±%) | L06[b] | (±%)[c] |
| $y^2D^o_{5/2}$ | 41761.24 | $a^2D_{3/2}$ | 0.00 | 2393.84 | 41761.24 | 0.254 | (5) | . . . | . . . |
| | | $a^2D_{5/2}$ | 3050.88 | 2582.52 | 38710.36 | 0.104 | (4) | . . . | . . . |
| | | $a^4F_{3/2}$ | 3644.65 | 2622.75 | 38116.59 | 0.204 | (3) | . . . | . . . |
| | | $a^4F_{5/2}$ | 4904.85 | 2712.43 | 36856.39 | 0.112 | (2) | . . . | . . . |
| | | $a^4F_{7/2}$ | 6344.34 | 2822.68 | 35416.90 | 0.231 | (2) | . . . | . . . |
| | | $a^2F_{5/2}$ | 12070.46 | 3367.08 | 29690.78 | 0.00207 | (14) | . . . | . . . |
| | | $a^2F_{7/2}$ | 15084.26 | 3747.49 | 26676.98 | 0.0152 | (10) | . . . | . . . |
| | | $b^2D_{3/2}$ | 14359.42 | 3648.35 | 27401.82 | 0.0122 | (8) | . . . | . . . |
| | | $a^2P_{3/2}$ | 17830.34 | 4177.52 | 23930.90 | 0.0379 | (14) | . . . | . . . |
| | | $b^4F_{5/2}$ | 20134.94 | 4622.70 | 21626.30 | 0.0252 | (16) | . . . | . . . |
| $z^4P^o_{5/2}$ | 40506.86 | $a^2D_{3/2}$ | 0.00 | 2467.97 | 40506.86 | 0.0907 | (5) | . . . | . . . |
| | | $a^2D_{5/2}$ | 3050.88 | 2669.01 | 37455.98 | 0.0709 | (4) | . . . | . . . |
| | | $a^4F_{3/2}$ | 3644.65 | 2712.00 | 36862.21 | 0.0350 | (3) | . . . | . . . |
| | | $a^4F_{5/2}$ | 4904.85 | 2808.00 | 35602.01 | 0.281 | (3) | . . . | . . . |
| | | $a^4F_{7/2}$ | 6344.34 | 2926.33 | 34162.52 | 0.0123 | (4) | . . . | . . . |
| | | $a^4P_{3/2}$ | 12920.94 | 3624.01 | 27585.92 | 0.108 | (8) | . . . | . . . |
| | | $a^4P_{5/2}$ | 13485.56 | 3699.73 | 27021.30 | 0.276 | (6) | . . . | . . . |
| | | $a^2F_{5/2}$ | 12070.46 | 3515.61 | 28436.40 | 0.00342 | (10) | . . . | . . . |
| | | $a^2F_{7/2}$ | 15084.26 | 3932.39 | 25422.60 | 0.0290 | (11) | . . . | . . . |
| | | $b^2D_{5/2}$ | 17368.87 | 4320.68 | 23137.99 | 0.0788 | (13) | . . . | . . . |
| | | $b^4F_{5/2}$ | 20134.94 | 4907.35 | 20371.92 | 0.00563 | (19) | . . . | . . . |
| $z^2F^o_{5/2}$ | 38578.63 | $a^2D_{3/2}$ | 0.00 | 2591.33 | 38578.63 | 0.0260 | (5) | . . . | . . . |
| | | $a^2D_{5/2}$ | 3050.88 | 2813.87 | 35527.75 | 0.0938 | (3) | . . . | . . . |
| | | $a^4F_{3/2}$ | 3644.65 | 2861.70 | 34933.98 | 0.427 | (1) | . . . | . . . |
| | | $a^4F_{5/2}$ | 4904.85 | 2968.80 | 33673.78 | 0.223 | (2) | . . . | . . . |
| | | $a^4F_{7/2}$ | 6344.34 | 3101.39 | 32234.29 | 0.145 | (2) | . . . | . . . |
| | | $a^4P_{5/2}$ | 13485.56 | 3984.04 | 25093.07 | 0.0117 | (10) | . . . | . . . |
| | | $a^2F_{5/2}$ | 12070.46 | 3771.35 | 26508.17 | 0.0117 | (9) | . . . | . . . |
| | | $b^2D_{3/2}$ | 14359.42 | 4127.79 | 24219.21 | 0.0321 | (10) | . . . | . . . |
| | | $a^2G_{7/2}$ | 17710.72 | 4790.71 | 20867.91 | 0.0127 | (15) | . . . | . . . |
| | | $b^4F_{3/2}$ | 18897.64 | 5079.63 | 19680.99 | 0.0156 | (15) | . . . | . . . |
| $z^2F^o_{7/2}$ | 41406.86 | $a^2D_{5/2}$ | 3050.88 | 2606.38 | 38355.98 | 0.207 | (2) | . . . | . . . |

| | | | | | | | | |
|---|---|---|---|---|---|---|---|---|
| | | $a^4F_{5/2}$ | 4904.85 | 2738.77 | 36502.01 | 0.391 | (1) | . . . | . . . |
| | | $a^4F_{7/2}$ | 6344.34 | 2851.21 | 35062.52 | 0.178 | (2) | . . . | . . . |
| | | $a^4F_{9/2}$ | 8361.76 | 3025.29 | 33045.10 | 0.0645 | (4) | . . . | . . . |
| | | $a^4P_{5/2}$ | 13485.56 | 3580.47 | 27921.30 | 0.00785 | (10) | . . . | . . . |
| | | $a^2F_{5/2}$ | 12070.46 | 3407.76 | 29336.40 | 0.0450 | (7) | . . . | . . . |
| | | $a^2F_{7/2}$ | 15084.26 | 3797.94 | 26322.60 | 0.0222 | (11) | . . . | . . . |
| | | $b^2D_{5/2}$ | 17368.87 | 4158.91 | 24037.99 | 0.0250 | (11) | . . . | . . . |
| | | $a^2G_{9/2}$ | 17389.06 | 4162.41 | 24017.80 | 0.0395 | (10) | . . . | . . . |
| | | $b^4F_{5/2}$ | 20134.94 | 4699.72 | 21271.92 | 0.0141 | (12) | . . . | . . . |
| | | $b^4F_{7/2}$ | 21637.97 | 5057.04 | 19768.89 | 0.00444 | (21) | . . . | . . . |
| $y^4F^o_{3/2}$ | 42518.10 | $a^2D_{3/2}$ | 0.00 | 2351.22 | 42518.10 | 0.332 | (2) | 0.323 | (17) |
| | | $a^2D_{5/2}$ | 3050.88 | 2532.99 | 39467.22 | 0.00697 | (2) | 0.006 | (16) |
| | | $a^4F_{3/2}$ | 3644.65 | 2571.68 | 38873.45 | 0.471 | (1) | 0.492 | (12) |
| | | $a^4F_{5/2}$ | 4904.85 | 2657.85 | 37613.25 | 0.0756 | (2) | 0.068 | (19) |
| | | $a^4P_{5/2}$ | 13485.56 | 3443.42 | 29032.54 | 0.00131 | (14) | 0.001 | (15) |
| | | $a^2F_{5/2}$ | 12070.46 | 3283.38 | 30447.64 | 0.0144 | (9) | 0.013 | (18) |
| | | $b^2D_{5/2}$ | 17368.87 | 3975.14 | 25149.23 | 0.00591 | (21) | 0.005 | (19) |
| | | $a^2P_{1/2}$ | 15254.29 | 3666.82 | 27263.81 | 0.00542 | (12) | 0.005 | (16) |
| | | $a^2P_{3/2}$ | 17830.34 | 4049.45 | 24687.76 | 0.0215 | (13) | 0.018 | (17) |
| | | $b^4F_{3/2}$ | 18897.64 | 4232.43 | 23620.46 | 0.0497 | (11) | 0.050 | (17) |
| | | $b^4F_{5/2}$ | 20134.94 | 4466.39 | 22383.16 | 0.0167 | (14) | 0.018 | (17) |
| $y^4F^o_{5/2}$ | 43680.75 | $a^2D_{3/2}$ | 0.00 | 2288.63 | 43680.75 | 0.00281 | (5) | 0.002 | (44) |
| | | $a^2D_{5/2}$ | 3050.88 | 2460.50 | 40629.87 | 0.322 | (2) | 0.337 | (14) |
| | | $a^4F_{3/2}$ | 3644.65 | 2496.99 | 40036.10 | 0.148 | (2) | 0.154 | (18) |
| | | $a^4F_{5/2}$ | 4904.85 | 2578.15 | 38775.90 | 0.129 | (2) | 0.133 | (16) |
| | | $a^4F_{7/2}$ | 6344.34 | 2677.56 | 37336.41 | 0.00712 | (8) | 0.007 | (16) |
| | | $a^4P_{3/2}$ | 12920.94 | 3250.06 | 30759.81 | 0.00237 | (12) | 0.001 | (13) |
| | | $a^4P_{5/2}$ | 13485.56 | 3310.83 | 30195.19 | 0.0182 | (10) | 0.018 | (17) |
| | | $a^2F_{5/2}$ | 12070.46 | 3162.61 | 31610.29 | 0.0910 | (10) | 0.062 | (51) |
| | | $a^2F_{7/2}$ | 15084.26 | 3495.93 | 28596.49 | 0.0665 | (11) | 0.066 | (16) |
| | | $b^2D_{5/2}$ | 17368.87 | 3799.49 | 26311.88 | 0.00778 | (13) | . . . | . . . |
| | | $a^2P_{3/2}$ | 17830.34 | 3867.31 | 25850.41 | 0.0412 | (10) | 0.042 | (15) |
| | | $a^2G_{7/2}$ | 17710.72 | 3849.50 | 25970.03 | 0.121 | (9) | 0.126 | (15) |
| | | $b^4F_{3/2}$ | 18897.64 | 4033.87 | 24783.11 | 0.0117 | (15) | 0.011 | (22) |
| | | $b^4F_{5/2}$ | 20134.94 | 4245.84 | 23545.81 | 0.0140 | (12) | 0.014 | (16) |
| | | $b^4F_{7/2}$ | 21637.97 | 4535.36 | 22042.78 | 0.0128 | (13) | 0.013 | (14) |
| | | $b^2G_{7/2}$ | 28458.16 | 6567.37 | 15222.59 | 0.00467 | (26) | 0.005 | (15) |
| $y^4F^o_{7/2}$ | 44399.96 | $a^2D_{5/2}$ | 3050.88 | 2417.70 | 41349.08 | 0.191 | (5) | . . . | . . . |
| | | $a^4F_{5/2}$ | 4904.85 | 2531.20 | 39495.11 | 0.174 | (4) | . . . | . . . |
| | | $a^4F_{7/2}$ | 6344.34 | 2626.95 | 38055.62 | 0.0308 | (3) | . . . | . . . |
| | | $a^4F_{9/2}$ | 8361.76 | 2774.01 | 36038.20 | 0.137 | (3) | . . . | . . . |
| | | $a^4P_{5/2}$ | 13485.56 | 3233.81 | 30914.40 | 0.00832 | (6) | . . . | . . . |
| | | $a^2F_{5/2}$ | 12070.46 | 3092.25 | 32329.50 | 0.0407 | (4) | . . . | . . . |
| | | $a^2F_{7/2}$ | 15084.26 | 3410.16 | 29315.70 | 0.143 | (6) | . . . | . . . |
| | | $b^2D_{5/2}$ | 17368.87 | 3698.39 | 27031.09 | 0.0475 | (9) | . . . | . . . |
| | | $a^2G_{9/2}$ | 17389.06 | 3701.16 | 27010.90 | 0.206 | (7) | . . . | . . . |
| | | $a^2G_{7/2}$ | 17710.72 | 3745.76 | 26689.24 | 0.00312 | (10) | . . . | . . . |
| | | $b^4F_{9/2}$ | 23145.57 | 4703.59 | 21254.39 | 0.00990 | (14) | . . . | . . . |
| | | $b^2G_{9/2}$ | 28104.83 | 6135.10 | 16295.13 | 0.00896 | (25) | . . . | . . . |
| $y^4F^o_{9/2}$ | 46124.89 | $a^4F_{7/2}$ | 6344.34 | 2513.04 | 39780.55 | 0.302 | (2) | . . . | . . . |
| | | $a^4F_{9/2}$ | 8361.76 | 2647.30 | 37763.13 | 0.433 | (2) | . . . | . . . |
| | | $a^2F_{7/2}$ | 15084.26 | 3220.65 | 31040.63 | 0.0796 | (6) | . . . | . . . |
| | | $a^2G_{9/2}$ | 17389.06 | 3478.98 | 28735.83 | 0.107 | (8) | . . . | . . . |
| | | $a^2G_{7/2}$ | 17710.72 | 3518.37 | 28414.17 | 0.00536 | (22) | . . . | . . . |
| | | $b^4F_{9/2}$ | 23145.57 | 4350.52 | 22979.32 | 0.0671 | (9) | . . . | . . . |
| | | $b^2G_{7/2}$ | 28458.16 | 5658.79 | 17666.73 | 0.000530 | (24) | . . . | . . . |
| | | $a^2H_{11/2}$ | 30941.97 | 6584.53 | 15182.92 | 0.00292 | (22) | . . . | . . . |
| $y^2P^o_{1/2}$ | 43044.26 | $a^2D_{3/2}$ | 0.00 | 2322.48 | 43044.26 | 0.785 | (1) | 0.806 | (5) |
| | | $a^4F_{3/2}$ | 3644.65 | 2537.33 | 39399.61 | 0.0844 | (4) | 0.083 | (21) |
| | | $a^4P_{3/2}$ | 12920.94 | 3318.73 | 30123.32 | 0.00347 | (16) | 0.003 | (20) |
| | | $b^2D_{3/2}$ | 14359.42 | 3485.16 | 28684.84 | 0.0141 | (14) | 0.010 | (19) |
| | | $a^2P_{1/2}$ | 15254.29 | 3597.39 | 27789.97 | 0.0563 | (15) | 0.049 | (19) |
| | | $a^2P_{3/2}$ | 17830.34 | 3964.94 | 25213.92 | 0.0408 | (16) | 0.039 | (22) |
| $y^2F^o_{5/2}$ | 43900.56 | $a^2D_{3/2}$ | 0.00 | 2277.17 | 43900.56 | 0.281 | (2) | . . . | . . . |
| | | $a^2D_{5/2}$ | 3050.88 | 2447.26 | 40849.68 | 0.215 | (2) | . . . | . . . |
| | | $a^4F_{3/2}$ | 3644.65 | 2483.36 | 40255.91 | 0.0196 | (2) | . . . | . . . |
| | | $a^4F_{5/2}$ | 4904.85 | 2563.62 | 38995.71 | 0.125 | (2) | . . . | . . . |
| | | $a^4F_{7/2}$ | 6344.34 | 2661.88 | 37556.22 | 0.140 | (3) | . . . | . . . |
| | | $a^4P_{3/2}$ | 12920.94 | 3227.00 | 30979.62 | 0.00733 | (10) | . . . | . . . |
| | | $a^2F_{5/2}$ | 12070.46 | 3140.77 | 31830.10 | 0.0373 | (9) | . . . | . . . |
| | | $a^2F_{7/2}$ | 15084.26 | 3469.27 | 28816.30 | 0.00246 | (13) | . . . | . . . |
| | | $b^2D_{3/2}$ | 14359.42 | 3384.14 | 29541.14 | 0.0427 | (11) | . . . | . . . |
| | | $a^2G_{7/2}$ | 17710.72 | 3817.19 | 26189.84 | 0.0548 | (9) | . . . | . . . |
| | | $b^4F_{3/2}$ | 18897.64 | 3998.40 | 25002.92 | 0.00702 | (15) | . . . | . . . |
| | | $b^4F_{5/2}$ | 20134.94 | 4206.57 | 23765.62 | 0.0568 | (10) | . . . | . . . |
| | | $b^4F_{7/2}$ | 21637.97 | 4490.58 | 22262.59 | 0.00988 | (15) | . . . | . . . |
| $y^2F^o_{7/2}$ | 44690.72 | $a^2D_{5/2}$ | 3050.88 | 2400.82 | 41639.84 | 0.0394 | (3) | . . . | . . . |
| | | $a^4F_{5/2}$ | 4904.85 | 2512.70 | 39785.87 | 0.264 | (1) | . . . | . . . |
| | | $a^4F_{7/2}$ | 6344.34 | 2607.03 | 38346.38 | 0.360 | (1) | . . . | . . . |
| | | $a^4F_{9/2}$ | 8361.76 | 2751.81 | 36328.96 | 0.115 | (2) | . . . | . . . |
| | | $a^4P_{5/2}$ | 13485.56 | 3203.67 | 31205.16 | 0.0330 | (9) | . . . | . . . |
| | | $a^2F_{5/2}$ | 12070.46 | 3064.69 | 32620.26 | 0.0338 | (6) | . . . | . . . |
| | | $a^2F_{7/2}$ | 15084.26 | 3376.67 | 29606.46 | 0.00429 | (10) | . . . | . . . |
| | | $b^2D_{5/2}$ | 17368.87 | 3659.03 | 27321.85 | 0.0108 | (11) | . . . | . . . |
| | | $a^2G_{9/2}$ | 17389.06 | 3661.74 | 27301.66 | 0.00754 | (23) | . . . | . . . |
| | | $a^2G_{7/2}$ | 17710.72 | 3705.40 | 26980.00 | 0.0447 | (11) | . . . | . . . |
| | | $b^4F_{5/2}$ | 20134.94 | 4071.21 | 24555.78 | 0.00577 | (15) | . . . | . . . |
| | | $b^4F_{7/2}$ | 21637.97 | 4336.66 | 23052.75 | 0.0643 | (10) | . . . | . . . |
| | | $b^4F_{9/2}$ | 23145.57 | 4640.12 | 21545.15 | 0.0126 | (22) | . . . | . . . |
| | | $b^2G_{9/2}$ | 28104.83 | 6027.55 | 16585.89 | 0.00421 | (20) | . . . | . . . |
| $y^4D^o_{3/2}$ | 46674.36 | $a^2D_{3/2}$ | 0.00 | 2141.83 | 46674.36 | 0.0121 | (5) | 0.010 | (15) |
| | | $a^2D_{5/2}$ | 3050.88 | 2291.64 | 43623.48 | 0.0197 | (2) | 0.020 | (16) |
| | | $a^4F_{3/2}$ | 3644.65 | 2323.26 | 43029.71 | 0.187 | (1) | 0.183 | (16) |
| | | $a^4F_{5/2}$ | 4904.85 | 2393.36 | 41769.51 | 0.459 | (1) | 0.447 | (10) |
| | | $a^4P_{1/2}$ | 11951.70 | 2879.12 | 34722.66 | 0.0331 | (9) | 0.033 | (18) |
| | | $a^4P_{3/2}$ | 12920.94 | 2961.80 | 33753.42 | 0.0779 | (7) | 0.077 | (18) |
| | | $a^4P_{5/2}$ | 13485.56 | 3012.19 | 33188.80 | 0.00856 | (12) | 0.008 | (14) |
| | | $a^2F_{5/2}$ | 12070.46 | 2889.00 | 34603.90 | 0.0108 | (9) | 0.010 | (14) |
| | | $a^2P_{1/2}$ | 15254.29 | 3181.76 | 31420.07 | 0.00888 | (11) | 0.008 | (14) |
| | | $b^4F_{3/2}$ | 18897.64 | 3599.11 | 27776.72 | 0.0292 | (13) | 0.028 | (15) |
| | | $b^4F_{5/2}$ | 20134.94 | 3766.91 | 26539.42 | 0.142 | (11) | 0.154 | (10) |
| | | $b^4P_{3/2}$ | 27285.13 | 5156.07 | 19389.23 | 0.00513 | (27) | 0.003 | (13) |
| $y^4D^o_{7/2}$ | 48930.75 | $a^2D_{5/2}$ | 3050.88 | 2178.92 | 45879.87 | 0.119 | (5) | . . . | . . . |
| | | $a^4F_{5/2}$ | 4904.85 | 2270.69 | 44025.90 | 0.00262 | (13) | . . . | . . . |
| | | $a^4F_{7/2}$ | 6344.34 | 2347.45 | 42586.41 | 0.201 | (3) | . . . | . . . |
| | | $a^4F_{9/2}$ | 8361.76 | 2464.19 | 40568.99 | 0.472 | (2) | . . . | . . . |
| | | $a^4P_{5/2}$ | 13485.56 | 2820.43 | 35445.19 | 0.0301 | (6) | . . . | . . . |
| | | $a^2F_{5/2}$ | 12070.46 | 2712.14 | 36860.29 | 0.0241 | (4) | . . . | . . . |
| | | $a^2F_{7/2}$ | 15084.26 | 2953.65 | 33846.49 | 0.00364 | (8) | . . . | . . . |
| | | $b^2D_{5/2}$ | 17368.87 | 3167.46 | 31561.88 | 0.00371 | (20) | . . . | . . . |
| | | $a^2G_{7/2}$ | 17710.72 | 3202.15 | 31220.03 | 0.0113 | (10) | . . . | . . . |
| | | $b^4F_{9/2}$ | 23145.57 | 3877.10 | 25785.18 | 0.113 | (13) | . . . | . . . |
| | | $b^4P_{5/2}$ | 28547.05 | 4904.51 | 20383.70 | 0.0183 | (14) | . . . | . . . |
| | | $b^2G_{7/2}$ | 28458.16 | 4883.22 | 20472.59 | 0.00145 | (21) | . . . | . . . |
| $z^2G^o_{7/2}$ | 47157.57 | $a^2D_{5/2}$ | 3050.88 | 2266.53 | 44106.69 | 0.0357 | (9) | . . . | . . . |
| | | $a^4F_{5/2}$ | 4904.85 | 2365.99 | 42252.72 | 0.0148 | (8) | . . . | . . . |
| | | $a^4F_{7/2}$ | 6344.34 | 2449.44 | 40813.23 | 0.0937 | (5) | . . . | . . . |
| | | $a^4F_{9/2}$ | 8361.76 | 2576.83 | 38795.81 | 0.223 | (3) | . . . | . . . |
| | | $a^4P_{5/2}$ | 13485.56 | 2968.96 | 33672.01 | 0.0317 | (3) | . . . | . . . |
| | | $a^2F_{5/2}$ | 12070.46 | 2849.21 | 35087.11 | 0.296 | (2) | . . . | . . . |
| | | $a^2F_{7/2}$ | 15084.26 | 3116.95 | 32073.31 | 0.0273 | (3) | . . . | . . . |
| | | $b^2D_{5/2}$ | 17368.87 | 3356.01 | 29788.70 | 0.00704 | (6) | . . . | . . . |
| | | $a^2G_{9/2}$ | 17389.06 | 3358.29 | 29768.51 | 0.0331 | (6) | . . . | . . . |
| | | $a^2G_{7/2}$ | 17710.72 | 3394.98 | 29446.85 | 0.129 | (5) | . . . | . . . |
| | | $b^4F_{5/2}$ | 20134.94 | 3699.55 | 27022.63 | 0.0108 | (11) | . . . | . . . |
| | | $b^4F_{7/2}$ | 21637.97 | 3917.45 | 25519.60 | 0.0461 | (11) | . . . | . . . |
| | | $b^4F_{9/2}$ | 23145.57 | 4163.41 | 24012.00 | 0.00833 | (20) | . . . | . . . |
| | | $b^4P_{5/2}$ | 28547.05 | 5371.81 | 18610.52 | 0.00256 | (21) | . . . | . . . |
| | | $b^2G_{9/2}$ | 28104.83 | 5247.13 | 19052.74 | 0.0231 | (16) | . . . | . . . |
| | | $b^2G_{7/2}$ | 28458.16 | 5346.28 | 18699.41 | 0.0173 | (17) | . . . | . . . |
| $x^2D^o_{3/2}$ | 47973.56 | $a^2D_{3/2}$ | 0.00 | 2083.82 | 47973.56 | 0.0847 | (7) | . . . | . . . |
| | | $a^4F_{3/2}$ | 3644.65 | 2255.17 | 44328.91 | 0.103 | (5) | . . . | . . . |
| | | $a^4F_{5/2}$ | 4904.85 | 2321.16 | 43068.71 | 0.203 | (5) | . . . | . . . |
| | | $a^4P_{1/2}$ | 11951.70 | 2775.27 | 36021.86 | 0.0550 | (9) | . . . | . . . |
| | | $a^4P_{3/2}$ | 12920.94 | 2852.02 | 35052.62 | 0.188 | (8) | . . . | . . . |
| | | $a^4P_{5/2}$ | 13485.56 | 2898.71 | 34488.00 | 0.238 | (15) | . . . | . . . |
| | | $a^2F_{5/2}$ | 12070.46 | 2784.45 | 35903.10 | 0.00295 | (9) | . . . | . . . |
| | | $b^2D_{5/2}$ | 17368.87 | 3266.53 | 30604.69 | 0.00759 | (14) | . . . | . . . |
| | | $a^2P_{1/2}$ | 15254.29 | 3055.42 | 32719.27 | 0.0626 | (11) | . . . | . . . |
| | | $b^4F_{3/2}$ | 18897.64 | 3438.29 | 29075.92 | 0.00929 | (16) | . . . | . . . |
| | | $b^4F_{5/2}$ | 20134.94 | 3591.11 | 27838.62 | 0.00463 | (18) | . . . | . . . |
| | | $b^4P_{1/2}$ | 26996.51 | 4765.78 | 20977.05 | 0.0120 | (21) | . . . | . . . |
| $x^2D^o_{5/2}$ | 49005.64 | $a^2D_{5/2}$ | 3050.88 | 2175.37 | 45954.76 | 0.0654 | (10) | . . . | . . . |
| | | $a^4F_{5/2}$ | 4904.85 | 2266.83 | 44100.79 | 0.160 | (7) | . . . | . . . |
| | | $a^4F_{7/2}$ | 6344.34 | 2343.33 | 42661.30 | 0.186 | (6) | . . . | . . . |
| | | $a^4P_{3/2}$ | 12920.94 | 2770.44 | 36084.70 | 0.0647 | (3) | . . . | . . . |
| | | $a^4P_{5/2}$ | 13485.56 | 2814.48 | 35520.08 | 0.270 | (2) | . . . | . . . |
| | | $a^2F_{5/2}$ | 12070.46 | 2706.64 | 36935.18 | 0.0681 | (3) | . . . | . . . |
| | | $a^2F_{7/2}$ | 15084.26 | 2947.13 | 33921.38 | 0.0495 | (3) | . . . | . . . |
| | | $b^2D_{3/2}$ | 14359.42 | 2885.47 | 34646.22 | 0.0510 | (2) | . . . | . . . |
| | | $a^2P_{3/2}$ | 17830.34 | 3206.74 | 31175.30 | 0.0151 | (5) | . . . | . . . |
| | | $a^2G_{7/2}$ | 17710.72 | 3194.48 | 31294.92 | 0.0258 | (5) | . . . | . . . |
| | | $b^4P_{3/2}$ | 27285.13 | 4602.65 | 21720.51 | 0.0102 | (20) | . . . | . . . |
| | | $b^2G_{7/2}$ | 28458.16 | 4865.42 | 20547.48 | 0.0159 | (23) | . . . | . . . |
| | | $c^2D_{3/2}$ | 30594.63 | 5430.02 | 18411.01 | 0.00266 | (27) | . . . | . . . |
| $w^2F^o_{7/2}$ | 53227.27 | $a^4F_{5/2}$ | 4904.85 | 2068.77 | 48322.42 | 0.00558 | (14) | . . . | . . . |
| | | $a^4F_{7/2}$ | 6344.34 | 2132.30 | 46882.93 | 0.00696 | (12) | . . . | . . . |
| | | $a^4F_{9/2}$ | 8361.76 | 2228.19 | 44865.51 | 0.0124 | (17) | . . . | . . . |
| | | $a^4P_{5/2}$ | 13485.56 | 2515.49 | 39741.71 | 0.187 | (4) | . . . | . . . |
| | | $a^2F_{5/2}$ | 12070.46 | 2428.99 | 41156.81 | 0.136 | (5) | . . . | . . . |
| | | $a^2F_{7/2}$ | 15084.26 | 2620.93 | 38143.01 | 0.0106 | (5) | . . . | . . . |
| | | $b^2D_{5/2}$ | 17368.87 | 2787.92 | 35858.40 | 0.00366 | (4) | . . . | . . . |
| | | $a^2G_{9/2}$ | 17389.06 | 2789.50 | 35838.21 | 0.184 | (2) | . . . | . . . |
| | | $a^2G_{7/2}$ | 17710.72 | 2814.76 | 35516.55 | 0.199 | (2) | . . . | . . . |
| | | $b^4F_{9/2}$ | 23145.57 | 3323.32 | 30081.70 | 0.0880 | (5) | . . . | . . . |
| | | $b^2G_{9/2}$ | 28104.83 | 3979.38 | 25122.44 | 0.124 | (14) | . . . | . . . |
| | | $a^2H_{9/2}$ | 31877.74 | 4682.63 | 21349.53 | 0.00776 | (22) | . . . | . . . |

Notes:

[a] Upper and lower levels are taken from NIST ASD and are ordered by term.

[b] L06: Lundqvist et al. (2006)

[c] Uncertainties are those quoted in L06 for *gf* values reduced in quadrature by their lifetime uncertainties.

Table 3. *A*-values and log(*gf*)s for 199 transitions of Hf II

| $\lambda_{air}$ (Å) | $E_{upper}$ ($cm^{-1}$) | $J_{upper}$ | $E_{lower}$ ($cm^{-1}$) | $J_{lower}$ | $A_{ki}$ ($10^6$ $s^{-1}$) | $\Delta A_{ki}$ ($10^6$ $s^{-1}$) | log(*gf*) |
|---|---|---|---|---|---|---|---|
| 2068.77 | 53227.27 | 3.5 | 4904.85 | 2.5 | 2.2 | 0.4 | -1.95 |
| 2083.82 | 47973.56 | 1.5 | 0.00 | 1.5 | 34. | 3. | -1.05 |
| 2132.30 | 53227.27 | 3.5 | 6344.34 | 3.5 | 2.7 | 0.4 | -1.83 |
| 2141.83 | 46674.36 | 1.5 | 0.00 | 1.5 | 6.1 | 0.7 | -1.78 |
| 2175.37 | 49005.64 | 2.5 | 3050.88 | 2.5 | 25. | 3. | -0.97 |
| 2178.92 | 48930.75 | 3.5 | 3050.88 | 2.5 | 57. | 6. | -0.49 |

Note: The table is available in its entirety in the online version. Only a portion of the table is shown here to indicate its form and content.

Table 4. Hf Abundances in HD 196944 and HD 222925

| | | | | HD 196944 | | HD 222925 | |
|---|---|---|---|---|---|---|---|
| $\lambda_{air}$ (Å) | $E_{lower}$ (eV) | log(*gf*) | Ref. | log $\varepsilon$ | $\sigma_{fit}$ | log $\varepsilon$ | $\sigma_{fit}$ |
| 2277.17 | 0.000 | -0.31 | 1 | -0.81 | 0.15 | 0.23 | 0.20 |
| 2322.48 | 0.000 | -0.14 | 1 | -0.64 | 0.15 | 0.09 | 0.10 |
| 2323.26 | 0.452 | -0.52 | 1 | -0.88 | 0.25 | 0.26 | 0.15 |
| 2393.36 | 0.608 | -0.10 | 1 | ... | ... | 0.15 | 0.20 |
| 2449.44 | 0.787 | -0.68 | 1 | ... | ... | 0.41 | 0.20 |
| 2622.75 | 0.452 | -0.36 | 1 | -0.59 | 0.15 | ... | ... |
| 2641.41 | 1.037 | +0.57 | 2 | -0.85 | 0.15 | 0.31 | 0.15 |
| 2647.30 | 1.037 | +0.21 | 1 | -0.65 | 0.10 | 0.43 | 0.10 |
| 2661.88 | 0.787 | -0.48 | 1 | ... | ... | 0.41 | 0.15 |
| 2738.77 | 0.608 | -0.17 | 1 | -0.78 | 0.25 | 0.25 | 0.20 |
| 2820.23 | 0.378 | -0.05 | 2 | -0.75 | 0.15 | ... | ... |
| 2861.02 | 0.000 | -0.69 | 2 | ... | ... | 0.38 | 0.20 |
| 2861.70 | 0.452 | -0.22 | 1 | -0.80 | 0.10 | 0.17 | 0.15 |
| 2919.60 | 0.452 | -0.71 | 2 | ... | ... | 0.47 | 0.15 |
| 2968.80 | 0.608 | -0.47 | 1 | ... | ... | 0.31 | 0.25 |
| 3012.90 | 0.000 | -0.60 | 2 | -0.69 | 0.05 | 0.36 | 0.10 |
| 3109.11 | 0.787 | -0.26 | 2 | ... | ... | 0.43 | 0.20 |
| 3399.79 | 0.000 | -0.57 | 2 | ... | ... | 0.25 | 0.10 |
| 3505.22 | 1.036 | -0.14 | 2 | ... | ... | 0.46 | 0.20 |
| 3719.28 | 0.608 | -0.81 | 2 | ... | ... | 0.34 | 0.15 |
| 3918.09 | 0.452 | -1.14 | 2 | -0.68 | 0.20 | 0.48 | 0.15 |
| 4093.15 | 0.452 | -1.15 | 2 | -0.73 | 0.15 | 0.34 | 0.05 |

**References**—1: this study; 2: UW07

Title: IMPROVED ATOMIC TRANSITION PROBABILITIES FOR UV AND OPTICAL LINES OF Hf II
Authors: E. A. Den Hartog, J. E. Lawler and I. U. Roederer
Table 3. A-values and log(gf)s for 199 transitions of Hf II.

Byte-by-byte Description of file: Table3_mr.txt

| Bytes | Format | Units | Label | Explanations |
|---|---|---|---|---|
| 1- 7 | F7.2 | Angstrom | Wave | Wavelength in air |
| 9- 16 | F8.2 | cm-1 | UpLev | Upper level |
| 18- 20 | F3.1 | none | UpJ | Upper level J value |
| 22- 29 | F8.2 | cm-1 | LowLev | Lower level |
| 31- 33 | F3.1 | none | LowJ | Lower level J value |
| 35- 40 | F6.2 | 10+6/s | TranP | Transition probability |
| 42- 46 | F5.2 | 10+6/s | e_TranP | Total uncertainty in TranP |
| 48- 52 | F5.2 | none | log(gf) | Log of the degeneracy times the oscillator strength |

| | | | | | | | |
|---|---|---|---|---|---|---|---|
| 2068.77 | 53227.27 | 3.5 | 4904.85 | 2.5 | 2.2 | 0.4 | -1.95 |
| 2083.82 | 47973.56 | 1.5 | 0.00 | 1.5 | 34. | 3. | -1.05 |
| 2132.30 | 53227.27 | 3.5 | 6344.34 | 3.5 | 2.7 | 0.4 | -1.83 |
| 2141.83 | 46674.36 | 1.5 | 0.00 | 1.5 | 6.1 | 0.7 | -1.78 |
| 2175.37 | 49005.64 | 2.5 | 3050.88 | 2.5 | 25. | 3. | -0.97 |
| 2178.92 | 48930.75 | 3.5 | 3050.88 | 2.5 | 57. | 6. | -0.49 |
| 2228.19 | 53227.27 | 3.5 | 8361.76 | 4.5 | 4.9 | 0.9 | -1.54 |
| 2255.17 | 47973.56 | 1.5 | 3644.65 | 1.5 | 41. | 4. | -0.90 |
| 2266.53 | 47157.57 | 3.5 | 3050.88 | 2.5 | 11.2 | 1.2 | -1.16 |
| 2266.83 | 49005.64 | 2.5 | 4904.85 | 2.5 | 62. | 6. | -0.55 |
| 2270.69 | 48930.75 | 3.5 | 4904.85 | 2.5 | 1.25 | 0.20 | -2.11 |
| 2277.17 | 43900.56 | 2.5 | 0.00 | 1.5 | 104. | 8. | -0.31 |
| 2288.63 | 43680.75 | 2.5 | 0.00 | 1.5 | 0.88 | 0.07 | -2.38 |
| 2291.64 | 46674.36 | 1.5 | 3050.88 | 2.5 | 9.9 | 1.0 | -1.51 |
| 2321.16 | 47973.56 | 1.5 | 4904.85 | 2.5 | 81. | 8. | -0.58 |
| 2322.48 | 43044.26 | 0.5 | 0.00 | 1.5 | 450 | 50 | -0.14 |
| 2323.26 | 46674.36 | 1.5 | 3644.65 | 1.5 | 93. | 9. | -0.52 |
| 2343.33 | 49005.64 | 2.5 | 6344.34 | 3.5 | 72. | 7. | -0.45 |
| 2347.45 | 48930.75 | 3.5 | 6344.34 | 3.5 | 96. | 10. | -0.20 |
| 2351.22 | 42518.10 | 1.5 | 0.00 | 1.5 | 154. | 15. | -0.29 |
| 2365.99 | 47157.57 | 3.5 | 4904.85 | 2.5 | 4.6 | 0.5 | -1.51 |
| 2393.36 | 46674.36 | 1.5 | 4904.85 | 2.5 | 230. | 23. | -0.10 |
| 2393.84 | 41761.24 | 2.5 | 0.00 | 1.5 | 88. | 7. | -0.35 |
| 2400.82 | 44690.72 | 3.5 | 3050.88 | 2.5 | 14.6 | 1.2 | -1.00 |
| 2417.70 | 44399.96 | 3.5 | 3050.88 | 2.5 | 47. | 3. | -0.49 |
| 2428.99 | 53227.27 | 3.5 | 12070.46 | 2.5 | 53. | 5. | -0.42 |
| 2447.26 | 43900.56 | 2.5 | 3050.88 | 2.5 | 80. | 6. | -0.37 |
| 2449.44 | 47157.57 | 3.5 | 6344.34 | 3.5 | 29.3 | 2.4 | -0.68 |
| 2460.50 | 43680.75 | 2.5 | 3050.88 | 2.5 | 101. | 7. | -0.26 |
| 2464.19 | 48930.75 | 3.5 | 8361.76 | 4.5 | 225. | 22. | 0.21 |
| 2467.97 | 40506.86 | 2.5 | 0.00 | 1.5 | 7.3 | 0.5 | -1.40 |
| 2483.36 | 43900.56 | 2.5 | 3644.65 | 1.5 | 7.3 | 0.6 | -1.39 |
| 2496.99 | 43680.75 | 2.5 | 3644.65 | 1.5 | 46. | 3. | -0.59 |
| 2512.70 | 44690.72 | 3.5 | 4904.85 | 2.5 | 98. | 7. | -0.13 |
| 2513.04 | 46124.89 | 4.5 | 6344.34 | 3.5 | 108. | 8. | 0.01 |
| 2515.49 | 53227.27 | 3.5 | 13485.56 | 2.5 | 73. | 6. | -0.25 |
| 2531.20 | 44399.96 | 3.5 | 4904.85 | 2.5 | 42.4 | 2.7 | -0.49 |
| 2532.99 | 42518.10 | 1.5 | 3050.88 | 2.5 | 3.2 | 0.3 | -1.90 |
| 2537.33 | 43044.26 | 0.5 | 3644.65 | 1.5 | 48. | 6. | -1.03 |
| 2563.62 | 43900.56 | 2.5 | 4904.85 | 2.5 | 46. | 4. | -0.56 |
| 2571.68 | 42518.10 | 1.5 | 3644.65 | 1.5 | 219. | 21. | -0.06 |
| 2576.83 | 47157.57 | 3.5 | 8361.76 | 4.5 | 70. | 5. | -0.26 |
| 2578.15 | 43680.75 | 2.5 | 4904.85 | 2.5 | 40.3 | 2.7 | -0.62 |
| 2582.52 | 41761.24 | 2.5 | 3050.88 | 2.5 | 35.7 | 2.9 | -0.67 |
| 2591.33 | 38578.63 | 2.5 | 0.00 | 1.5 | 5.0 | 0.3 | -1.52 |
| 2606.38 | 41406.86 | 3.5 | 3050.88 | 2.5 | 39.9 | 2.2 | -0.49 |
| 2607.03 | 44690.72 | 3.5 | 6344.34 | 3.5 | 134. | 10. | 0.04 |
| 2620.93 | 53227.27 | 3.5 | 15084.26 | 3.5 | 4.2 | 0.4 | -1.46 |
| 2622.75 | 41761.24 | 2.5 | 3644.65 | 1.5 | 70. | 5. | -0.36 |
| 2626.95 | 44399.96 | 3.5 | 6344.34 | 3.5 | 7.5 | 0.4 | -1.21 |
| 2647.30 | 46124.89 | 4.5 | 8361.76 | 4.5 | 155. | 11. | 0.21 |
| 2657.85 | 42518.10 | 1.5 | 4904.85 | 2.5 | 35. | 3. | -0.83 |
| 2661.88 | 43900.56 | 2.5 | 6344.34 | 3.5 | 52. | 4. | -0.48 |
| 2669.01 | 40506.86 | 2.5 | 3050.88 | 2.5 | 5.7 | 0.4 | -1.44 |
| 2677.56 | 43680.75 | 2.5 | 6344.34 | 3.5 | 2.23 | 0.23 | -1.84 |
| 2706.64 | 49005.64 | 2.5 | 12070.46 | 2.5 | 26.2 | 2.2 | -0.76 |
| 2712.00 | 40506.86 | 2.5 | 3644.65 | 1.5 | 2.82 | 0.17 | -1.73 |
| 2712.14 | 48930.75 | 3.5 | 12070.46 | 2.5 | 11.5 | 1.2 | -1.00 |
| 2712.43 | 41761.24 | 2.5 | 4904.85 | 2.5 | 39. | 3. | -0.59 |
| 2738.77 | 41406.86 | 3.5 | 4904.85 | 2.5 | 75. | 4. | -0.17 |
| 2751.81 | 44690.72 | 3.5 | 8361.76 | 4.5 | 43. | 3. | -0.41 |
| 2770.44 | 49005.64 | 2.5 | 12920.94 | 1.5 | 24.9 | 2.1 | -0.76 |
| 2774.01 | 44399.96 | 3.5 | 8361.76 | 4.5 | 33.4 | 1.9 | -0.51 |
| 2775.27 | 47973.56 | 1.5 | 11951.70 | 0.5 | 22.0 | 2.7 | -0.99 |
| 2784.45 | 47973.56 | 1.5 | 12070.46 | 2.5 | 1.18 | 0.14 | -2.26 |
| 2787.92 | 53227.27 | 3.5 | 17368.87 | 2.5 | 1.44 | 0.13 | -1.87 |
| 2789.50 | 53227.27 | 3.5 | 17389.06 | 4.5 | 72. | 6. | -0.17 |
| 2808.00 | 40506.86 | 2.5 | 4904.85 | 2.5 | 22.6 | 1.4 | -0.79 |
| 2813.87 | 38578.63 | 2.5 | 3050.88 | 2.5 | 18.0 | 1.1 | -0.89 |
| 2814.48 | 49005.64 | 2.5 | 13485.56 | 2.5 | 104. | 8. | -0.13 |
| 2814.76 | 53227.27 | 3.5 | 17710.72 | 3.5 | 78. | 6. | -0.13 |
| 2820.43 | 48930.75 | 3.5 | 13485.56 | 2.5 | 14.4 | 1.6 | -0.86 |
| 2822.68 | 41761.24 | 2.5 | 6344.34 | 3.5 | 80. | 6. | -0.24 |
| 2849.21 | 47157.57 | 3.5 | 12070.46 | 2.5 | 93. | 6. | -0.04 |
| 2851.21 | 41406.86 | 3.5 | 6344.34 | 3.5 | 34.3 | 1.9 | -0.48 |
| 2852.02 | 47973.56 | 1.5 | 12920.94 | 1.5 | 75. | 9. | -0.43 |
| 2861.70 | 38578.63 | 2.5 | 3644.65 | 1.5 | 82. | 4. | -0.22 |
| 2879.12 | 46674.36 | 1.5 | 11951.70 | 0.5 | 16.6 | 2.2 | -1.08 |
| 2885.47 | 49005.64 | 2.5 | 14359.42 | 1.5 | 19.6 | 1.6 | -0.83 |
| 2889.00 | 46674.36 | 1.5 | 12070.46 | 2.5 | 5.4 | 0.7 | -1.57 |
| 2898.71 | 47973.56 | 1.5 | 13485.56 | 2.5 | 95. | 16. | -0.32 |
| 2926.33 | 40506.86 | 2.5 | 6344.34 | 3.5 | 0.99 | 0.06 | -2.12 |
| 2947.13 | 49005.64 | 2.5 | 15084.26 | 3.5 | 19.0 | 1.6 | -0.83 |
| 2953.65 | 48930.75 | 3.5 | 15084.26 | 3.5 | 1.73 | 0.22 | -1.74 |
| 2961.80 | 46674.36 | 1.5 | 12920.94 | 1.5 | 39. | 5. | -0.69 |
| 2968.80 | 38578.63 | 2.5 | 4904.85 | 2.5 | 42.8 | 2.3 | -0.47 |
| 2968.96 | 47157.57 | 3.5 | 13485.56 | 2.5 | 9.9 | 0.7 | -0.98 |
| 3012.19 | 46674.36 | 1.5 | 13485.56 | 2.5 | 4.3 | 0.7 | -1.63 |
| 3025.29 | 41406.86 | 3.5 | 8361.76 | 4.5 | 12.4 | 0.8 | -0.87 |
| 3055.42 | 47973.56 | 1.5 | 15254.29 | 0.5 | 25. | 3. | -0.85 |
| 3064.69 | 44690.72 | 3.5 | 12070.46 | 2.5 | 12.5 | 1.2 | -0.85 |
| 3092.25 | 44399.96 | 3.5 | 12070.46 | 2.5 | 9.9 | 0.6 | -0.94 |
| 3101.39 | 38578.63 | 2.5 | 6344.34 | 3.5 | 27.9 | 1.5 | -0.62 |
| 3116.95 | 47157.57 | 3.5 | 15084.26 | 3.5 | 8.5 | 0.6 | -1.00 |
| 3140.77 | 43900.56 | 2.5 | 12070.46 | 2.5 | 13.8 | 1.6 | -0.91 |
| 3162.61 | 43680.75 | 2.5 | 12070.46 | 2.5 | 28. | 3. | -0.59 |
| 3167.46 | 48930.75 | 3.5 | 17368.87 | 2.5 | 1.8 | 0.4 | -1.67 |
| 3181.76 | 46674.36 | 1.5 | 15254.29 | 0.5 | 4.4 | 0.7 | -1.57 |
| 3194.48 | 49005.64 | 2.5 | 17710.72 | 3.5 | 9.9 | 0.9 | -1.04 |
| 3202.15 | 48930.75 | 3.5 | 17710.72 | 3.5 | 5.4 | 0.7 | -1.18 |
| 3203.67 | 44690.72 | 3.5 | 13485.56 | 2.5 | 12.2 | 1.4 | -0.82 |
| 3206.74 | 49005.64 | 2.5 | 17830.34 | 1.5 | 5.8 | 0.5 | -1.27 |
| 3220.65 | 46124.89 | 4.5 | 15084.26 | 3.5 | 28.4 | 2.7 | -0.35 |
| 3227.00 | 43900.56 | 2.5 | 12920.94 | 1.5 | 2.7 | 0.3 | -1.59 |
| 3233.81 | 44399.96 | 3.5 | 13485.56 | 2.5 | 2.03 | 0.16 | -1.59 |
| 3250.06 | 43680.75 | 2.5 | 12920.94 | 1.5 | 0.74 | 0.10 | -2.15 |
| 3266.53 | 47973.56 | 1.5 | 17368.87 | 2.5 | 3.0 | 0.5 | -1.71 |
| 3283.38 | 42518.10 | 1.5 | 12070.46 | 2.5 | 6.7 | 0.9 | -1.36 |
| 3310.83 | 43680.75 | 2.5 | 13485.56 | 2.5 | 5.7 | 0.7 | -1.25 |
| 3318.73 | 43044.26 | 0.5 | 12920.94 | 1.5 | 2.0 | 0.4 | -2.18 |
| 3323.32 | 53227.27 | 3.5 | 23145.57 | 4.5 | 35. | 3. | -0.34 |
| 3356.01 | 47157.57 | 3.5 | 17368.87 | 2.5 | 2.20 | 0.19 | -1.53 |
| 3358.29 | 47157.57 | 3.5 | 17389.06 | 4.5 | 10.4 | 0.9 | -0.85 |
| 3367.08 | 41761.24 | 2.5 | 12070.46 | 2.5 | 0.72 | 0.11 | -2.14 |
| 3376.67 | 44690.72 | 3.5 | 15084.26 | 3.5 | 1.59 | 0.20 | -1.66 |
| 3384.14 | 43900.56 | 2.5 | 14359.42 | 1.5 | 15.8 | 2.1 | -0.79 |
| 3394.98 | 47157.57 | 3.5 | 17710.72 | 3.5 | 40. | 3. | -0.25 |
| 3407.76 | 41406.86 | 3.5 | 12070.46 | 2.5 | 8.6 | 0.8 | -0.92 |
| 3410.16 | 44399.96 | 3.5 | 15084.26 | 3.5 | 34.8 | 2.7 | -0.31 |
| 3438.29 | 47973.56 | 1.5 | 18897.64 | 1.5 | 3.7 | 0.7 | -1.58 |
| 3443.42 | 42518.10 | 1.5 | 13485.56 | 2.5 | 0.61 | 0.10 | -2.36 |
| 3469.27 | 43900.56 | 2.5 | 15084.26 | 3.5 | 0.91 | 0.14 | -2.00 |
| 3478.98 | 46124.89 | 4.5 | 17389.06 | 4.5 | 38. | 4. | -0.16 |
| 3485.16 | 43044.26 | 0.5 | 14359.42 | 1.5 | 8.1 | 1.5 | -1.53 |
| 3495.93 | 43680.75 | 2.5 | 15084.26 | 3.5 | 20.8 | 2.6 | -0.64 |
| 3515.61 | 40506.86 | 2.5 | 12070.46 | 2.5 | 0.28 | 0.03 | -2.51 |
| 3518.37 | 46124.89 | 4.5 | 17710.72 | 3.5 | 1.9 | 0.4 | -1.45 |
| 3580.47 | 41406.86 | 3.5 | 13485.56 | 2.5 | 1.51 | 0.17 | -1.63 |
| 3591.11 | 47973.56 | 1.5 | 20134.94 | 2.5 | 1.9 | 0.4 | -1.84 |
| 3597.39 | 43044.26 | 0.5 | 15254.29 | 0.5 | 32. | 6. | -0.90 |
| 3599.11 | 46674.36 | 1.5 | 18897.64 | 1.5 | 14.6 | 2.4 | -0.95 |
| 3624.01 | 40506.86 | 2.5 | 12920.94 | 1.5 | 8.8 | 0.8 | -0.99 |
| 3648.35 | 41761.24 | 2.5 | 14359.42 | 1.5 | 4.2 | 0.4 | -1.30 |
| 3659.03 | 44690.72 | 3.5 | 17368.87 | 2.5 | 4.0 | 0.5 | -1.19 |
| 3661.74 | 44690.72 | 3.5 | 17389.06 | 4.5 | 2.8 | 0.7 | -1.35 |
| 3666.82 | 42518.10 | 1.5 | 15254.29 | 0.5 | 2.5 | 0.4 | -1.69 |
| 3698.39 | 44399.96 | 3.5 | 17368.87 | 2.5 | 11.6 | 1.2 | -0.72 |
| 3699.55 | 47157.57 | 3.5 | 20134.94 | 2.5 | 3.4 | 0.4 | -1.26 |
| 3699.73 | 40506.86 | 2.5 | 13485.56 | 2.5 | 22.2 | 1.8 | -0.56 |
| 3701.16 | 44399.96 | 3.5 | 17389.06 | 4.5 | 50. | 4. | -0.08 |
| 3705.40 | 44690.72 | 3.5 | 17710.72 | 3.5 | 16.6 | 2.2 | -0.56 |
| 3745.76 | 44399.96 | 3.5 | 17710.72 | 3.5 | 0.76 | 0.08 | -1.89 |
| 3747.49 | 41761.24 | 2.5 | 15084.26 | 3.5 | 5.2 | 0.6 | -1.18 |
| 3766.91 | 46674.36 | 1.5 | 20134.94 | 2.5 | 71. | 11. | -0.22 |
| 3771.35 | 38578.63 | 2.5 | 12070.46 | 2.5 | 2.25 | 0.24 | -1.54 |
| 3797.94 | 41406.86 | 3.5 | 15084.26 | 3.5 | 4.3 | 0.5 | -1.13 |
| 3799.49 | 43680.75 | 2.5 | 17368.87 | 2.5 | 2.4 | 0.4 | -1.50 |
| 3817.19 | 43900.56 | 2.5 | 17710.72 | 3.5 | 20.3 | 2.4 | -0.58 |
| 3849.50 | 43680.75 | 2.5 | 17710.72 | 3.5 | 38. | 4. | -0.30 |
| 3867.31 | 43680.75 | 2.5 | 17830.34 | 1.5 | 12.9 | 1.5 | -0.76 |
| 3877.10 | 48930.75 | 3.5 | 23145.57 | 4.5 | 54. | 9. | -0.01 |
| 3917.45 | 47157.57 | 3.5 | 21637.97 | 3.5 | 14.4 | 1.9 | -0.58 |
| 3932.39 | 40506.86 | 2.5 | 15084.26 | 3.5 | 2.34 | 0.29 | -1.49 |
| 3964.94 | 43044.26 | 0.5 | 17830.34 | 1.5 | 23. | 5. | -0.96 |
| 3975.14 | 42518.10 | 1.5 | 17368.87 | 2.5 | 2.8 | 0.6 | -1.58 |
| 3979.38 | 53227.27 | 3.5 | 28104.83 | 4.5 | 49. | 8. | -0.03 |
| 3984.04 | 38578.63 | 2.5 | 13485.56 | 2.5 | 2.25 | 0.26 | -1.49 |
| 3998.40 | 43900.56 | 2.5 | 18897.64 | 1.5 | 2.6 | 0.4 | -1.43 |
| 4033.87 | 43680.75 | 2.5 | 18897.64 | 1.5 | 3.6 | 0.6 | -1.27 |
| 4049.45 | 42518.10 | 1.5 | 17830.34 | 1.5 | 10.0 | 1.6 | -1.01 |
| 4071.21 | 44690.72 | 3.5 | 20134.94 | 2.5 | 2.1 | 0.4 | -1.37 |
| 4127.79 | 38578.63 | 2.5 | 14359.42 | 1.5 | 6.2 | 0.7 | -1.02 |
| 4158.91 | 41406.86 | 3.5 | 17368.87 | 2.5 | 4.8 | 0.6 | -1.00 |
| 4162.41 | 41406.86 | 3.5 | 17389.06 | 4.5 | 7.6 | 0.8 | -0.80 |
| 4163.41 | 47157.57 | 3.5 | 23145.57 | 4.5 | 2.6 | 0.6 | -1.27 |
| 4177.52 | 41761.24 | 2.5 | 17830.34 | 1.5 | 13.1 | 2.0 | -0.69 |
| 4206.57 | 43900.56 | 2.5 | 20134.94 | 2.5 | 21.1 | 2.6 | -0.47 |
| 4232.43 | 42518.10 | 1.5 | 18897.64 | 1.5 | 23. | 3. | -0.60 |
| 4245.84 | 43680.75 | 2.5 | 20134.94 | 2.5 | 4.4 | 0.6 | -1.15 |
| 4320.68 | 40506.86 | 2.5 | 17368.87 | 2.5 | 6.4 | 0.9 | -0.97 |
| 4336.66 | 44690.72 | 3.5 | 21637.97 | 3.5 | 23.8 | 2.9 | -0.27 |
| 4350.52 | 46124.89 | 4.5 | 23145.57 | 4.5 | 24.0 | 2.8 | -0.17 |
| 4466.39 | 42518.10 | 1.5 | 20134.94 | 2.5 | 7.8 | 1.3 | -1.03 |
| 4490.58 | 43900.56 | 2.5 | 21637.97 | 3.5 | 3.7 | 0.6 | -1.18 |
| 4535.36 | 43680.75 | 2.5 | 21637.97 | 3.5 | 4.0 | 0.6 | -1.13 |
| 4602.65 | 49005.64 | 2.5 | 27285.13 | 1.5 | 3.9 | 0.9 | -1.13 |
| 4622.70 | 41761.24 | 2.5 | 20134.94 | 2.5 | 8.7 | 1.5 | -0.78 |
| 4640.12 | 44690.72 | 3.5 | 23145.57 | 4.5 | 4.7 | 1.1 | -0.92 |
| 4682.63 | 53227.27 | 3.5 | 31877.74 | 4.5 | 3.0 | 0.7 | -1.10 |
| 4699.72 | 41406.86 | 3.5 | 20134.94 | 2.5 | 2.7 | 0.4 | -1.14 |
| 4703.59 | 44399.96 | 3.5 | 23145.57 | 4.5 | 2.4 | 0.4 | -1.19 |
| 4765.78 | 47973.56 | 1.5 | 26996.51 | 0.5 | 4.8 | 1.1 | -1.18 |
| 4790.71 | 38578.63 | 2.5 | 17710.72 | 3.5 | 2.4 | 0.4 | -1.30 |
| 4865.42 | 49005.64 | 2.5 | 28458.16 | 3.5 | 6.1 | 1.5 | -0.88 |
| 4883.22 | 48930.75 | 3.5 | 28458.16 | 3.5 | 0.69 | 0.16 | -1.70 |
| 4904.51 | 48930.75 | 3.5 | 28547.05 | 2.5 | 8.7 | 1.4 | -0.60 |
| 4907.35 | 40506.86 | 2.5 | 20134.94 | 2.5 | 0.45 | 0.09 | -2.01 |
| 5057.04 | 41406.86 | 3.5 | 21637.97 | 3.5 | 0.85 | 0.18 | -1.58 |
| 5079.63 | 38578.63 | 2.5 | 18897.64 | 1.5 | 3.0 | 0.5 | -1.16 |
| 5156.07 | 46674.36 | 1.5 | 27285.13 | 1.5 | 2.6 | 0.7 | -1.39 |
| 5247.13 | 47157.57 | 3.5 | 28104.83 | 4.5 | 7.2 | 1.3 | -0.62 |
| 5346.28 | 47157.57 | 3.5 | 28458.16 | 3.5 | 5.4 | 1.0 | -0.73 |
| 5371.81 | 47157.57 | 3.5 | 28547.05 | 2.5 | 0.80 | 0.18 | -1.56 |
| 5430.02 | 49005.64 | 2.5 | 30594.63 | 1.5 | 1.02 | 0.29 | -1.57 |
| 5658.79 | 46124.89 | 4.5 | 28458.16 | 3.5 | 0.19 | 0.05 | -2.04 |
| 6027.55 | 44690.72 | 3.5 | 28104.83 | 4.5 | 1.6 | 0.3 | -1.17 |
| 6135.10 | 44399.96 | 3.5 | 28104.83 | 4.5 | 2.2 | 0.6 | -1.01 |
| 6567.37 | 43680.75 | 2.5 | 28458.16 | 3.5 | 1.5 | 0.4 | -1.25 |
| 6584.53 | 46124.89 | 4.5 | 30941.97 | 5.5 | 1.04 | 0.24 | -1.17 |